\documentclass[11pt]{article}

\usepackage{amsmath,amssymb}           
\usepackage{amscd}                     
\usepackage{epsfig}                    
\usepackage[matrix,arrow]{xy}          
\usepackage{xspace}                    
\usepackage{stmaryrd}                  
\usepackage{slashed}

\usepackage{jheppub}                   

\DeclareSymbolFont{bbold}{U}{bbold}{m}{n}
\DeclareSymbolFontAlphabet{\mathbbold}{bbold}

\newcommand{\bq}{\begin{equation}}
\newcommand{\eq}{\end{equation}}
\newcommand{\bea}{\begin{eqnarray}}
\newcommand{\eea}{\end{eqnarray}}

\newcommand{\dd}{\mathrm{d}}
\newcommand{\ee}{\mathrm{e}}
\newcommand{\ii}{\mathrm{i}}

\newcommand{\der}{\partial}

\newcommand{\bbZ}{\mathbb{Z}}
\newcommand{\bbR}{\mathbb{R}}
\newcommand{\bbC}{\mathbb{C}}
\newcommand{\bbH}{\mathbb{H}}

\DeclareMathOperator{\SU}{\mathit{SU}}
\DeclareMathOperator{\SO}{\mathit{SO}}
\DeclareMathOperator{\SL}{\mathit{SL}}
\DeclareMathOperator{\GL}{\mathit{GL}}

\DeclareMathOperator{\Spin}{\mathit{Spin}}

\DeclareMathOperator{\gl}{\mathit{gl}}

\DeclareMathOperator{\Cliff}{Cliff}

\newcommand{\rep}[1]{\mathbf{#1}}
\newcommand{\repp}[2]{(\rep{#1}, \rep{#2})}
\newcommand{\id}{\mathbbold{1}}

\DeclareMathOperator{\vol}{vol}

\newcommand{\Gs}[1]{\Gamma(#1)}

\newcommand{\ph}[1]{\phantom{#1}}

\newcommand{\Lgen}{L}

\newcommand{\Bgen}[2]{\left\llbracket#1,#2\right\rrbracket}
\newcommand{\BLie}[2]{\left[#1,#2\right]}

\newcommand{\Dgen}{{D}}

\DeclareMathOperator{\adj}{ad}

\newcommand{\bl}[2]{\langle{#1},{#2}\rangle}

\newcommand{\volG}{\left|\vol_G\right|}

\newcommand{\LC}{\nabla}

\newcommand{\Riem}{\mathcal{R}}
\newcommand{\Ric}{\mathcal{R}}
\newcommand{\Scalar}{\mathcal{R}}

\newcommand{\GenRic}{R^{\scriptscriptstyle 0}}
\newcommand{\GenRicci}{R}
\newcommand{\GenS}{R}

\DeclareMathOperator{\Diff}{Diff}

\newcommand{\proj}[1]{\times_{#1}}

\newcommand{\inn}{\mathbin{\lrcorner}}
\newcommand{\oadj}{\proj{\text{ad}}}

\DeclareMathOperator{\Edd}{\mathit{E_{d(d)}}}
\DeclareMathOperator{\Hd}{\mathit{H_d}}
\DeclareMathOperator{\dHd}{\mathit{\tilde{H}_d}}
\DeclareMathOperator{\E7}{\mathit{E}_{7(7)}}

\newcommand{\tA}{{\tilde{A}}}
\newcommand{\tF}{{\tilde{F}}}
\newcommand{\ta}{{\tilde{a}}}
\newcommand{\tb}{{\tilde{b}}}
\newcommand{\talpha}{{\tilde{\alpha}}}
\newcommand{\tLambda}{{\tilde{\Lambda}}}
\newcommand{\tV}{\tilde{V}}

\newcommand{\am}{Q}
\newcommand{\dex}{c}
\newcommand{\Hperp}{\adj{P}^\perp}

\newcommand{\Vout}{\curlywedge}
\newcommand{\Vin}{\curlyvee}

\newcommand{\DSS}{\slashed{\Dgen}}
\newcommand{\DJJ}{\slashed{\Dgen}}
\newcommand{\DSJ}{\Dgen \Vout}
\newcommand{\DJS}{\Dgen \Vin}

\newcommand{\gamh}{\hat{\gamma}}

\title{Supergravity as Generalised Geometry II: \\ $\Edd\times\bbR^+$ and
  M theory} 

\author{Andr\'e Coimbra$^a$, 
  Charles Strickland-Constable$^{a,b}$  
  and Daniel Waldram$^a$}

\affiliation{$^a$Department of Physics,
   Imperial College London \\
   Prince Consort Road, London, SW7 2AZ, UK \\
   \\
   $^b$II. Institut f\"ur Theoretische Physik 
   der Universit\"at Hamburg, \\
   Luruper Chaussee 149, D-22761 Hamburg, Germany \\}

\emailAdd{a.coimbra08@imperial.ac.uk}
\emailAdd{charles.strickland.constable@desy.de}
\emailAdd{d.waldram@imperial.ac.uk}

\subheader{\textrm{Imperial/TP/12/DW/01}}

\abstract{
We reformulate eleven-dimensional supergravity, including fermions, in
terms of generalised geometry, for spacetimes that are warped
products of Minkowski space with a $d$-dimensional manifold $M$ with
$d\leq7$. The reformulation has a $E_{d(d)}\times\mathbb{R}^+$ structure group and
is has a local $\tilde{H}_d$ symmetry, where $\tilde{H}_d$ is the double cover of
the maximally compact subgroup of $E_{d(d)}$. The bosonic degrees for
freedom unify into a generalised metric, and, defining the generalised
analogue $D$ of the Levi--Civita connection, one finds that the
corresponding equations of motion are the vanishing of the generalised
Ricci tensor. To leading order, we show that the fermionic equations of
motion, action and supersymmetry variations can all be written in
terms of $D$. Although we will not give the detailed
decompositions, this reformulation is equally applicable to type IIA
or IIB supergravity restricted to a $(d-1)$-dimensional manifold. For
completeness we give explicit expressions in terms of
$\tilde{H}_4=Spin(5)$ and $\tilde{H}_7=SU(8)$ representations for
$d=4$ and $d=7$. 
}

\begin{document}
\maketitle


\section{Introduction}
\label{sec:intro}

In this paper we describe the reformulation eleven-dimensional
supergravity in terms of generalised geometry. The analysis is
restricted to spacetimes that are warped products of Minkowski space
with a $d$-dimensional manifold $M$ with $d\leq7$ and we work to
leading order in the fermions. The generalised geometry has a
$\Edd\times\bbR^+$ structure group and the theory has a local $\dHd$
symmetry, where $\dHd$ is the double cover of the maximally compact
subgroup of $\Edd$. Although we will not give the detailed
decompositions, this reformulation is equally applicable to type IIA
or IIB supergravity restricted to a $(d-1)$-dimensional manifold.   

This completes a programme started in~\cite{CSW1}, where we studied
the formulation of type II supergravity in terms of a generalised
geometry with an $O(d,d)\times\bbR^+$ structure group first proposed by
Hitchin and Gualtieri~\cite{GCY,Gualtieri}. Such reformulations were
first given in the related ``doubled'' formalism in a series of papers
by Hohm, Hull, Kwak and Zweibach~\cite{double1}, building in part on
work by Siegel~\cite{siegel}. This was extended to the
RR sector in~\cite{double2} (along with~\cite{CSW1}). In such
Double Field Theory, rather than extending only the tangent space as
in generalised geometry, one conjectures that the spacetime is doubled
and, to relate to supergravity, then imposes a constraint that there
is dependence on only half the coordinates. Interestingly, even when
applied to conventional supergravity backgrounds, the doubled space cannot be
assumed to have a conventional manifold structure~\cite{double-geom}.  

In~\cite{CSW1}, we focussed in particular on the notion of
\emph{generalised connections}. We showed that there exists a class of
torsion-free, metric-compatible generalised connections, analogues of
the Levi--Civita connection. (These objects, the corresponding
curvature tensors and the relation to the NSNS sector of supergravity,
were first discussed in the ``doubled'' formalism by
Siegel~\cite{siegel}. Related ``semi-covariant'' derivatives, and
the corresponding curvature tensors were defined independently by
Jeon, Lee and Park in~\cite{JLP}.) We then used these results to write
down a gravitational theory for generalised geometry and showed that
this was precisely type II supergravity with its local ``double
Lorentz symmetry'' $O(9,1)\times O(1,9)$ manifest. In the following
paper~\cite{CSW2}, we defined the analogous concepts 
in $\Edd\times\bbR^+$ generalised geometry, also known as exceptional
or extended generalised geometry~\cite{chris,PW}. We showed that this
construction can be used to describe eleven-dimensional supergravity
restricted to $d$ dimensions, that is, to the warped product of
Minkowski space and a $d$-dimensional manifold. The action and field
equations for the bosonic sector are simply the $\Edd\times\bbR^+$
generalised version of Einstein's gravity  
\begin{equation}
   S_{\text{B}} = \frac{1}{2\kappa^2}\int \volG \GenS \, , \qquad 
   \GenRicci_{AB} = 0 ,
\end{equation}
where $\volG$ is the volume density associated to the generalised
metric $G$, and $\GenRicci_{AB}$ and $\GenS$ are respectively the
generalised Ricci tensor and scalar associated to a torsion-free,
metric compatible generalised connection $\Dgen$.  

The present paper serves as a direct continuation of the
$\Edd\times\bbR^+$ construction, by including the fermion fields in
the description of the supergravity. We show how the same generalised
connection describes the supersymmetry algebra, and can be used to
obtain the fermion  dynamics. The equations of motion for the two
fermionic fields present in the theory, $\psi$ and $\rho$, are simply 
\begin{equation}
\begin{aligned}
\label{eq:intro-ferm-eoms}
   \DJJ \psi
      + \tfrac{11-d}{9-d}\DSJ \rho  &= 0, \\
   \DJS \psi  + \DSS \rho  &= 0,
\end{aligned}
\end{equation}
where $\DJJ{}$, $\DSJ{}$ and $\DJS{}$ are particular $\dHd$-covariant
operators built from the generalised connection that project, in the
first line, onto the $\psi$ representation, and in the second line,
onto the $\rho$ representation. The geometry therefore naturally
produces a supergravity theory with the larger symmetries manifest,
including the fermions to first order. 

While the formalism is powerful enough to describe the theory in
different dimensions in all generality, the fact that the structures
appearing in each case are so diverse forces us to introduce fairly
abstract notation. It can therefore be helpful to look at explicit
constructions. We work out two examples with the full local symmetry
manifest, the relatively simple $d=4$ case and the more interesting
$d=7$ case. For instance, in the latter case, when the local group is
$\SU(8)$ the projections in~\eqref{eq:intro-ferm-eoms} become 
\begin{equation}
\begin{aligned}
   -\tfrac{1}{12} \, 
     \epsilon_{\alpha\beta\gamma\delta\delta'\theta_1\theta_2\theta_3} 
     \Dgen^{\delta \delta'} \psi^{\theta_1 \theta_2 \theta_3}
	+ 2\bar\Dgen_{[\alpha \beta} \bar{\rho}_{\gamma]} &= 0 ,\\
	- \tfrac{1}{2} \bar\Dgen_{\beta \gamma} \psi^{\alpha \beta \gamma} 
	+ \Dgen^{\alpha \beta} \bar{\rho}_\beta  &= 0 ,
\end{aligned}
\end{equation}
which is a remarkably compact form compared with the usual
supergravity expressions. 

There are many precursors and related approaches to the geometrical
reformulation discussed
here~\cite{duff2,deWN,nic,west,KNS,dewitBPS,deWN-extra,E10,hillmann,BP},
as well as several, more recent, papers that have developed or applied
ideas of exceptional generalised
geometry~\cite{GLSW,E7-flux,%
GO1,Thompson:2011uw,BMJ,BP-alg,Cassani:2011fu,GO2,BMT,BCKT,GT}.
Let us comment briefly on a couple of related approaches of particular 
relevance here. In 1986 de Wit and Nicolai~\cite{deWN} already showed
that a local $\SU(8)$ symmetry (and a global $\E7$ symmetry on an
extension of the usual vielbein) could be realised directly in $d=11$
supergravity, including the linearised fermions, assuming only the
existence of a product structure on the tangent bundle
$\SO(3,1)\times\SO(7)\subset\SO(10,1)$. The current paper can 
be viewed as a geometrical framework from which to interpret 
these results. In the bosonic sector, a more recent approach is the 
work of Berman and Perry and collaborators~\cite{BP}, using the
M-theory extension of double field theory~\cite{duff1,dft}. These
authors were able to find a non-manifestly covariant form of the action
simply in terms of derivatives of the generalised metric $G$. Again
the work of~\cite{CSW2} and this paper puts this result on a covariant 
geometrical footing, demonstrating that the action is none other than
the generalised Ricci tensor. One might also expect a connection to
special cases of the much broader non-linear $E_{11}$ proposal of
West~\cite{west}, along the lines of the relation between Riemannian
geometry and the non-linear realisation of gravity due to Borisov and
Ogievetsky~\cite{BO}. In a remarkably detailed
construction~\cite{hillmann}, including the fermions and using West's
approach, Hillmann indeed considered a ``generalised $E_{7(7)}$ coset
dynamics'', on a sixty-dimensional spacetime~\cite{deWN-extra}. By
demanding that, upon truncating to $4+7$ dimensions, the theory
possess $\Diff(7)$ invariance, the author managed to show that the
construction reproduces the results of de Wit and
Nicolai~\cite{deWN}. In generalised geometry one does not increase the
dimension of the underlying manifold, so clearly the two formalisms
are technically distinct, and the precise relation between them is yet
to be investigated. 

The paper is thus structured as follows. In section~\ref{sec:sugra} we
give a quick review eleven-dimensional supergravity and its
restrictions to $d$ dimensions. In section~\ref{sec:EddGG} we provide
a very brief summary of the key points in~\cite{CSW2}. In
section~\ref{sec:GenGrav} we introduce the fermion fields and provide
a complete rewrite of supergravity in terms of the generalised
geometry formalism. In section~\ref{sec:explicit} we work out the
explicit $d=4,7$ cases. We conclude with a short discussion in
section~\ref{sec:conc}. We also include a number of appendices to fix
our conventions and also give the details of the various spinor
decompositions and group actions that we use.


\section{Eleven-dimensional supergravity and its restriction}
\label{sec:sugra}

\subsection{$N=1$, $D=11$ supergravity}
\label{sec:11sugra}

Let us start by reviewing the action, equations of motion and
supersymmetry variations of eleven-dimensional supergravity, to 
leading order in the fermions, following the conventions
of~\cite{GP} (see also appendices~\ref{app:conv-d}
and~\ref{app:spin-decomp}). 

The fields are simply 
\begin{equation}
  \{ g_{MN}, \mathcal{A}_{MNP}, \psi_M\} , 
\end{equation}
where $g_{MN}$ is the metric, $\mathcal{A}_{MNP}$ the three-form
potential and $\psi_M$ is the gravitino. The bosonic action is given
by  
\begin{equation}
\label{eq:NSaction}
   S_{\text{B}} = \frac{1}{2\kappa^2}\int \Big(
       \sqrt{-g} \,\Scalar - \tfrac{1}{2}\mathcal{F}\wedge *\mathcal{F}
          - \tfrac{1}{6}\mathcal{A}\wedge \mathcal{F}\wedge
          \mathcal{F} 
       \Big) ,
\end{equation}
where $\Scalar$ is the Ricci scalar and $\mathcal{F}=\dd
\mathcal{A}$. This leads to the equations of motion  
\begin{equation}
\label{eq:eom11}
\begin{aligned}
  \Ric_{MN} - \tfrac{1}{12} \left( 
           \mathcal{F}_{MP_1P_2P_3} 
           \mathcal{F}^{\ph{N}P_1P_2P_3}_N
		- \tfrac{1}{12} g_{MN} \mathcal{F}^2 \right)
      &= 0 , \\
   \dd * \mathcal{F} + \tfrac{1}{2} \mathcal{F}\wedge \mathcal{F} &= 0 , 
\end{aligned}
\end{equation}
where $\Ric_{MN}$ is the Ricci tensor. 

Taking $\Gamma^M$ to be the $\Cliff(10,1;\bbR)$ gamma matrices, the
fermionic action, to quadratic order in $\psi_M$, is given by 
\begin{equation}
\begin{aligned}
   S_{\text{F}} &= \frac{1}{\kappa^2} \int \sqrt{-g} 
       \Big( \bar\psi_M \Gamma^{MNP} \LC_N \psi_P 
           + \tfrac{1}{96} \mathcal{F}_{P_1\dots P_4}
              \bar\psi_M \Gamma^{MP_1\dots P_4N}\psi_N
       \\ & \qquad \qquad \qquad \qquad \qquad 
           + \tfrac{1}{8} \mathcal{F}_{P_1\dots P_4}
              \bar\psi^{P_1}\Gamma^{P_2P_3}\psi^{P_4} 
		\Big) ,
\end{aligned}
\end{equation}
the gravitino equation of motion is
\begin{equation}
  \Gamma^{MNP} \LC_N \psi_P + \tfrac{1}{96} \left(
      \Gamma^{MNP_1\dots P_4} \mathcal{F}_{P_1\dots P_4}
      + 12 \mathcal{F}^{MN}{}_{P_1P_2} 
         \Gamma^{P_1P_2} \right) \psi_N = 0 .
\end{equation}
The supersymmetry variations of the bosons are
\begin{equation}
\begin{aligned}
   \delta g_{MN} &= 2 \bar\varepsilon \Gamma_{(M} \psi_{N)} ,\\
   \delta \mathcal{A}_{MNP} 
      &= -3 \bar\varepsilon \Gamma_{[MN} \psi_{P]} ,
\end{aligned}
\end{equation}
while the supersymmetry variation of the gravitino is
\begin{equation}
  \delta \psi_M = \LC_M \varepsilon 
     + \tfrac{1}{288} \left(\Gamma_M{}^{N_1\dots N_4}
        - 8 \delta_M{}^{N_1} \Gamma^{N_2N_3N_4} 
        \right) \mathcal{F}_{N_1\dots N_4} \varepsilon ,
\end{equation}
where $\varepsilon$ is the supersymmetry parameter.


\subsection{Restricted action, equations of motion and supersymmetry}
\label{sec:eom}

We will be interested in ``restrictions'' of eleven-dimensional
supergravity where the spacetime is assumed to be a warped product
$\bbR^{10-d,1}\times M$ of Minkowski space with a $d$-dimensional
spin manifold $M$, with $d\leq 7$. The metric is taken to
have the form  
\begin{equation}
   \dd s_{11}^2 = \ee^{2\Delta}\dd s^2(\bbR^{10-d,1}) + \dd s_d^2(M) , 
\end{equation}
where $\dd s^2(\bbR^{10-d,1})$ is the flat metric on $\bbR^{10-d,1}$
and $\dd s_d^2(M)$ is a general metric on $M$. The warp factor
$\Delta$ and all the other fields are assumed to be independent of the
flat $\bbR^{10-d,1}$ space. In this sense we restrict the full
eleven-dimensional theory to $M$. We will split the eleven-dimensional
indices as external indices $ \mu = 0, 1, \dots , \dex-1$ and internal
indices $m=1,\dots,d$ where $\dex+d=11$.

In the restricted theory, the surviving fields include the obvious
internal components of the eleven-dimensional fields (namely the
metric $g$ and three-form $A$) as well as the warp factor $\Delta$. If
$d=7$, the eleven-dimensional Hodge dual of the 4-form $F$ can have a
purely internal 7-form component. This leads one to introduce, in
addition, a dual six-form potential $\tilde{A}$ on $M$ which is
related to the seven-form field strength $\tilde{F}$ by  
\begin{equation}
   \tilde{F} = \dd \tilde{A} - \tfrac12 A \wedge F .
\end{equation}
The Bianchi identities satisfied by $F=\dd A$ and $\tilde{F}$ are then 
\begin{equation}
\begin{aligned}
   \dd F &= 0 , \\
   \dd \tilde{F} + \tfrac12 F \wedge F &= 0 . 
\end{aligned}
\end{equation}
With these definitions one can see that $F$ and $\tF$ are related to
the components of the eleven dimensional 4-form field strength
$\mathcal{F}$ by 
\begin{equation}
  F_{m_1 \dots m_4} = \mathcal{F}_{m_1 \dots m_4} , 
   \qquad \qquad \qquad 
   \tF_{m_1 \dots m_7} 
       = \left(*\mathcal{F}\right)_{m_1 \dots m_7} , 
\end{equation}
where $*\mathcal{F}$ is the eleven-dimensional Hodge dual. The field
strengths $F$ and $\tF$ are invariant under the gauge transformations
of the potentials given by 
\begin{equation}
\label{eq:AtA-transf}
\begin{aligned}
   A' &= A + \dd\Lambda , \\
   \tA' &= \tA + \dd\tLambda
       - \tfrac{1}{2}\dd\Lambda\wedge A ,
\end{aligned}
\end{equation}
for some two-form $\Lambda$ and five-form $\tLambda$. There is an
intricate hierarchy of further coupled gauge transformations of
$\Lambda$ and $\tLambda$, discussed in more detail in~\cite{PW}
and~\cite{CSW2} and which formally defines a form of
``gerbe''~\cite{gerbes}. 

In order to diagonalise the kinetic terms in the fermionic Lagrangian,
one introduces the standard field redefinition of the external
components of the gravitino 
\begin{equation}
  \psi'_\mu = \psi_\mu + \tfrac{1}{c-2} \Gamma_\mu \Gamma^m \psi_m.
\end{equation}
We then denote its trace as
\begin{equation}
  \rho = \tfrac{c-2}{c} \Gamma^\mu \psi'_\mu ,
\end{equation}
and allow this to be non-zero and dependant on the internal
coordinates (this is the partner of the warp factor
$\Delta$). Although the restriction to $d$-dimensions breaks the
Lorentz symmetry to $\Spin(10-d,1)\times\Spin(d)\subset\Spin(10,1)$, 
we do not make an explicit decomposition of the spinor indices under 
$\Spin(10-d,1)\times\Spin(d)$. Instead we keep expressions in terms of
eleven-dimensional gamma matrices. This is helpful in what follows
since it allows us to treat all dimensions in a uniform  way.  

In summary, the surviving degrees of freedom after the restriction to
$d$ dimensions are 
\begin{equation}
  \{ g_{mn}, A_{mnp}, \tilde{A}_{m_1 \dots m_6}, 
      \Delta; \psi_m, \rho \} .  
\end{equation}
One can then define the internal space bosonic action
\begin{equation}
\label{eq:Boson-action}
  S_{\text{B}} = \frac{1}{2\kappa^2}\int 
       \sqrt{g} \; \ee^{\dex\Delta} \Big(
        \Scalar + \dex(\dex-1) (\der \Delta)^2
         - \tfrac12 \tfrac{1}{4!} F^2 
          - \tfrac12 \tfrac{1}{7!} \tilde{F}^2 \Big) ,
\end{equation}
where the associated equations of motion 
\begin{equation}
\label{eq:eom11d}
\begin{aligned}
  \Ric_{mn} - \dex \LC_m \LC_n \Delta 
       - \dex(\der_m \Delta)(\der_n \Delta)  
       - \tfrac12 \tfrac{1}{4!}\big( 
             4F_{m p_1 p_2 p_3} F_n{}^{p_1 p_2 p_3}
            - \tfrac{1}{3} g_{mn} F^2 \big) \quad & \\
       - \tfrac12 \tfrac{1}{7!}\big(  
             7\tilde{F}_{m p_1 \dots p_6}\tilde{F}_n{}^{p_1 \dots p_6}
             - \tfrac{2}{3} g_{mn} \tilde{F}^2 \big) &= 0 , \\*[3pt]
	\Scalar - 2(\dex-1) \LC^2 \Delta - \dex(\dex-1) (\der \Delta)^2 
                - \tfrac12 \tfrac{1}{4!} F^2 
		- \tfrac12 \tfrac{1}{7!} \tilde{F}^2 
                &= 0 , \\*[3pt]
	\dd *(\ee^{\dex \Delta}F)
            - \ee^{\dex \Delta} F \wedge *\tilde{F} &= 0, \\*[3pt]
	\dd * (\ee^{\dex \Delta} \tilde{F}) &= 0,
\end{aligned}
\end{equation}
are those obtained by substituting the field ansatz
into~\eqref{eq:eom11}. Similarly, to quadratic order in fermions, the
action for the fermion fields is 
\begin{equation}
\label{eq:Fermion-action}
\begin{split}
   S_{\text{F}} = - & \frac{1}{\kappa^2(c-2)^2} 
      \int \sqrt{g} \, \ee^{c\Delta} \bigg[
        (c-4) \bar\psi_m \Gamma^{mnp} \LC_n \psi_p \\ & \qquad
        - c(c-3) \bar\psi^m \Gamma^n \LC_n \psi_m 
        - c \left( \bar\psi^m \Gamma_n \LC_m \psi^n 
           + \bar\psi^m \Gamma_m \LC_n \psi^n\right) 
        \\*[3pt] & \qquad\quad
        - \tfrac{1}{4} \tfrac{1}{2!} (2c^2 - 5c + 4) 
	   \bar\psi_m F^{mn}{}_{pq} \Gamma^{pq} \psi_n 
        + \tfrac{1}{4} c(c-3) \bar\psi_m \slashed{F} \psi^m 
        \\*[3pt] & \qquad\quad
        + \tfrac{1}{2} \tfrac{1}{3!}c \bar\psi_m
           F^{m}{}_{pqr} \Gamma^{npqr} \psi_n  
        + \tfrac{1}{4} \tfrac{1}{4!} (c-4)
           \bar\psi_m  F_{p_1 \dots p_4} \Gamma^{mnp_1\dots p_4}\psi_n 
        \\*[3pt] & \qquad\quad
        - \tfrac{1}{4} \tfrac{1}{5!} (2c^2 -5c +4) 
           \bar\psi_m \tilde{F}^{mn}{}_{p_1 \dots p_5}
              \Gamma^{p_1 \dots p_5} \psi_n 
        \\*[3pt] & \qquad\quad
        + \tfrac{1}{4} \tfrac{1}{6!} c(c-1) 
           \bar\psi_m \tilde{F}^m{}_{p_1 \dots p_6} 
              \Gamma^{np_1 \dots p_6} \psi_n 
        \\*[3pt] & \qquad
	+ c(c-1) \big(\bar\psi^m \LC_m \rho - \bar\rho \LC^m \psi_m \big)
        + c\big(\bar\psi_m \Gamma^{mn} \LC_n \rho  
           - \bar\rho \Gamma^{mn} \LC_m \psi_n\big)
        \\*[3pt] & \qquad\quad
	- c(c-1)(c-2) \bar\psi^m (\der_m \Delta) \rho 
        - c(c-2) \bar\psi_m \Gamma^{mn} (\der_n\Delta) \rho 
        \\*[3pt] & \qquad\quad
        + \tfrac{1}{2} \tfrac{1}{3!} c(c-1) \bar\rho
          F^m{}_{pqr} \Gamma^{pqr} \psi_m
        - \tfrac{1}{2} \tfrac{1}{4!} c \bar\rho 
           \Gamma^m{}_{p_1 \dots p_4} F^{p_1 \dots p_4} \psi_m
        \\*[3pt] & \qquad\quad
        - \tfrac{1}{2} \tfrac{1}{6!} c(c-1) 
           \bar\psi_m \tilde{F}^m{}_{p_1 \dots p_6} 
              \Gamma^{p_1 \dots p_6} \rho
        \\ & \qquad
        +  c(c-1) \big( \bar\rho \Gamma^m \LC_m \rho 
           + \tfrac{1}{4} \bar\rho \slashed{F} \rho
           - \tfrac{1}{4} \bar\rho \slashed{\tilde{F}} \rho \big) 
       \bigg] .
\end{split}
\end{equation}
This action leads to the equation of motion for $\psi_m$,
\begin{equation}
\label{eq:eom-psi}
\begin{split}
   0 &= (c-4) \Gamma_m{}^{np}\left( 
      \LC_n + \tfrac{c}{2} \der_n \Delta\right) \psi_p 
   - c(c-3) \Gamma^n \left( \LC_n + \tfrac{c}{2} 
      \der_n \Delta\right) \psi_m 
   \\ & \qquad
   - c \Gamma_n \left( \LC_m + \tfrac{c}{2} \der_m \Delta\right) \psi^n 
   - c \Gamma_m \left( \LC_n + \tfrac{c}{2} \der_n \Delta\right)\psi^n) 
   \\ & \qquad
   - \tfrac{1}{4} (2c^2 - 5c + 4) \tfrac{1}{2!} 
      F_{mnpq} \Gamma^{pq}\psi^n 
   + \tfrac{1}{4} c(c-3) \slashed{F} \psi_m 
   \\ & \qquad
   + \tfrac{1}{2}\tfrac{1}{3!}c F_{(m}{}^{p_1 p_2 p_3} 
      \Gamma_{n) p_1 p_2 p_3} \psi^n
   + \tfrac{1}{4}\tfrac{1}{4!} (c-4) \Gamma_{mn}{}^{p_1 \dots p_4} 
      F_{p_1 \dots p_4} \psi^n 
   \\ & \qquad
   - \tfrac{1}{4}  \tfrac{1}{5!} (2c^2 - 5c +4)
      \tilde{F}_{mnp_1 \dots p_5} 
      \Gamma^{p_1 \dots p_5} \psi^n
   + \tfrac{1}{4} \tfrac{1}{6!} c(c-1) \tilde{F}_{(m}{}^{p_1 \dots p_6} 
      \Gamma_{n)} {}_{p_1 \dots p_6} \psi^n
   \\ & \quad
   + c \Gamma_m{}^n \left(\LC_n+\der_n \Delta\right) \rho
      + c(c-1) \left(\LC_m +\der_m \Delta\right) \rho
   \\ & \qquad
   + \tfrac{1}{4} \tfrac{1}{3!} c(c-1) F_{m p_1 p_2 p_3} 
      \Gamma^{p_1 p_2 p_3} \rho
   + \tfrac{1}{4}  \tfrac{1}{4!} c \Gamma_{m p_1 \dots p_4} 
      F^{p_1\dots p_4} \rho 
   \\ & \qquad
   - \tfrac{1}{4} \tfrac{1}{6!} c(c-1) \tilde{F}_{m n_1 \dots n_6} 
      \Gamma^{n_1 \dots n_6} \rho , 
\end{split}
\end{equation}
and the equation of motion for $\rho$,
\begin{equation}
\label{eq:eom-rho}
\begin{split}
   0 &= \big[
         \slashed{\LC} + \tfrac{c}{2} (\slashed\der \Delta)
         + \tfrac{1}{4} \slashed{F} - \tfrac{1}{4} \slashed{\tilde{F}}
         \big] \rho 
   \\ & \qquad
   - \big[\LC_m + (c-1) \der_m \Delta\big] \psi^m 
   - \tfrac{1}{c-1} \Gamma^{mn} \big[\LC_m 
      + (c-1) \der_m \Delta\big] \psi_n
   \\ & \qquad
   + \tfrac{1}{4} \tfrac{1}{3!} F^m{}_{p_1 p_2 p_3} 
      \Gamma^{p_1 p_2 p_3} \psi_m 
   - \tfrac{1}{4} \tfrac{1}{4!} \tfrac{1}{c-1} 
      \Gamma^m{}_{p_1 \dots p_4} F^{p_1 \dots p_4} \psi_m
   \\ & \qquad
   + \tfrac{1}{4} \tfrac{1}{6!} \tilde{F}^m{}_{p_1 \dots p_6} 
      \Gamma^{p_1 \dots p_6} \psi_m . 
\end{split}
\end{equation}

Turning to the supersymmetry transformations, we find that the
variations of the fermion fields are given by 
\begin{equation}
\label{eq:susy-ferm}
\begin{split}
   \delta \rho &= \left[ 
      \slashed{\LC} - \tfrac{1}{4} \slashed{F} 
      - \tfrac{1}{4} \slashed{\tilde{F}} 
      + \tfrac{c-2}{2} (\slashed{\der} \Delta) 
      \right] \varepsilon ,\\
   \delta \psi_m &= \left[ 
      \LC_m + \tfrac{1}{288} F_{n_1 \dots n_4} \left(
         \Gamma_m{}^{n_1 \dots n_4} 
         - 8 \delta_{m}{}^{n_1} \Gamma^{n_2 n_3 n_4}\right)  
      - \tfrac{1}{12} \tfrac{1}{6!} \tilde{F}_{mn_1 \dots n_6} 
	 \Gamma^{n_1 \dots n_6} 
      \right] \varepsilon,
\end{split}
\end{equation}
and the variations of the bosons by
\begin{equation}
\label{eq:susy-bos}
\begin{split}
	\delta g_{mn} &= 2 \bar\varepsilon \Gamma_{(m} \psi_{n)} ,\\
	(c-2)\delta \Delta +  \delta \; \mathrm{log} \sqrt{g} &= \bar\varepsilon \rho ,\\
	\delta A_{mnp} &= - 3  \bar\varepsilon \Gamma_{[mn} \psi_{p]} ,\\
	\delta \tA_{m_1 \dots m_6} &= 6  \bar\varepsilon \Gamma_{[m_1 \dots m_5} \psi_{m_6]} .
\end{split}
\end{equation}
\setlength{\jot}{5pt}
This completes our summary of the reduced theory. 

In what follows the fermionic fields will be reinterpreted as
representations of larger symmetry groups $\dHd\supset\Spin(d)$. To
mark that distinction, the fermions that have appeared in this section
will be denoted by $\varepsilon^{\text{sugra}}$, $\rho^{\text{sugra}}$
and $\psi^{\text{sugra}}$. Absent this label, the fields are to be viewed
as ``generalised'' objects transforming under $\dHd$, as will be
clarified in section~\ref{sec:Hd}.


\section{Review of $\Edd\times\bbR^+$ generalised geometry}
\label{sec:EddGG}

We now give a brief summary of the key points in the construction
of the $\Edd\times\bbR^+$ generalised geometry and connections,
relevant to reductions of eleven-dimensional supergravity, as
discussed in~\cite{CSW2}. 


\subsection{Generalised bundles and frames}
\label{sec:gen-bundles}


Let $M$ be a $d$-dimensional spin manifold with $d\leq 7$\footnote{We
   actually only consider $4\leq d\leq 7$, as for lower dimensions the
   relevant structures simplify to a point that generalised geometry
   has little to add to the usual Riemannian description.}. The 
generalised tangent space $E$
is isomorphic to the sum~\cite{PW,chris} 
\begin{equation}
\label{eq:Eiso}
   E \simeq TM \oplus \Lambda^2T^*M \oplus \Lambda^5T^*M
         \oplus (T^*M \otimes\Lambda^7T^*M) ,
\end{equation}
where for $d<7$ some of these terms will of course be
absent. Physically the terms in the sum can be thought of as
corresponding to different brane charges, namely, momentum,
M2-brane, M5-brane and Kaluza--Klein monopole charge.  The bundle is
actually given by a series of extensions which are defined via the 
patching data of the three-form and six-form
connections. (Specifying particular three- and six-form connections
defines an isomorphism~\eqref{eq:Eiso}.) The patching
structure matches the supergravity symmetries~\eqref{eq:AtA-transf}
(see~\cite{PW,chris}). In this way the bundle $E$ encodes all the
topological information of the supergravity background: the twisting
of the tangent space $TM$ as well as that of the form-field potentials.

The fibre $E_x$ of the generalised vector bundle at $x\in M$ forms a
representation space of $\Edd\times\bbR^+$. These are listed in
Table~\ref{tab:gen-tang}. 
\begin{table}[htb]
\begin{center}
\begin{tabular}{lll}
   $\Edd$ group & $E$ rep. & $\adj{\tilde{F}}$ rep. \\
   \hline
   $E_{7(7)}$ & $\rep{56}_\rep{1}$ 
     & $\rep{133}_\rep{0} + \rep{1}_\rep{0}$ \\
   $E_{6(6)}$ & $\rep{27}'_\rep{1}$ 
     & $\rep{78}_\rep{0} + \rep{1}_\rep{0}$ \\
   $E_{5(5)}\simeq\Spin(5,5)$  & $\rep{16}^c_\rep{1}$ 
     & $\rep{45}_\rep{0} + \rep{1}_\rep{0}$ \\
   $E_{4(4)}\simeq\SL(5,\bbR)$ & $\rep{10}'_\rep{1}$ %
     & $\rep{24}_\rep{0} + \rep{1}_\rep{0}$ 
\end{tabular}
\end{center}
\caption{Generalised tangent space and frame bundle representations
   where the subscript denotes the $\bbR^+$ weight, where
   $\rep{1}_\rep{1}\simeq(\det{T^*M})^{1/(9-d)}$} 
\label{tab:gen-tang}
\end{table}
The definition of the $\Edd\times\bbR^+$ group and its explicit action
on $E_x$ is given in appendix~\ref{app:Edd}. Crucially, the patching
used to define $E$ is compatible with the $\Edd\times\bbR^+$
action. This means that one can define a generalised structure bundle
as a sub-bundle of the general frame bundle $F$ for $E$. Let
$\{\hat{E}_A\}$ be a basis for $E_x$, where the label $A$ runs over
the dimension of  the generalised tangent space as listed in
table~\ref{tab:gen-tang}. As usual, a choice of coordinates on a patch $U$
defines a particular such basis where 
\begin{equation}
\label{eq:coord}
   \{\hat{E}_A\}
      = \{\der/\der x^m\} 
      \cup\{\dd x^m\wedge\dd x^n\}
      \cup \{\dd x^{m_1}\wedge\dots\wedge\dd x^{m_5} \} 
      \cup \{ \dd x^m \otimes \dd x^{m_1}\wedge\dots\wedge\dd x^{m_7} 
      \}.
\end{equation}
We will denote the components of a generalised vector $V$ in such a
coordinate frame by an index $M$, namely $V^M = (v^m,\omega_{mn},
\sigma_{m_1\dots m_5},\tau_{m,m_1\dots m_7})$. 

The \emph{generalised structure bundle} is then the unique
$\Edd\times\bbR^+$ principle sub-bundle $\tilde{F} \subset F$
compatible with the patching. Concretely, it can be written as      
\begin{equation}
\label{eq:gen-fb}
   \tilde{F} = \big\{ (x,\{\hat{E}_A\}) : \text{$x\in M$, and 
        $\{\hat{E}_A\}$  is a $\Edd\times\bbR^+$ basis of $\tilde{E}_x$} 
      \big\} ,
\end{equation}
where an $\Edd\times\bbR^+$ basis is any choice of frame that is
related to the coordinate frame by an $\Edd\times\bbR^+$
transformation as defined in appendix~\ref{app:Edd}. By construction, this is a
principle bundle with fibre $\Edd\times\bbR^+$.

A special class of $\Edd\times\bbR^+$ frames are those defined by a
splitting of the generalised tangent space $E$, that is, an 
isomorphism of the form~\eqref{eq:Eiso}. As we mentioned, this is
equivalent to introducing the three- and six-form gauge potentials, $A$
and $\tA$.  Then, given a generic basis $\{\hat{e}_a\}$ for $TM$,
$\{e^a\}$ as the dual basis on $T^*M$ and a scalar function $\Delta$,
one has that a \emph{conformal split frame} $\{\hat{E}_A\}$ for
$\tilde{E}$ has the general form (see appendix~\ref{app:Edd} for
notation)  
\begin{equation}
\label{eq:geom-basis}
\begin{aligned}
   \hat{E}_a &= \ee^{\Delta} \Big( \hat{e}_a + i_{\hat{e}_a} A
      + i_{\hat{e}_a}\tA 
      + \tfrac{1}{2}A\wedge i_{\hat{e}_a}A 
      \\ & \qquad \qquad 
      + jA\wedge i_{\hat{e}_a}\tA 
      + \tfrac{1}{6}jA\wedge A \wedge i_{\hat{e}_a}A \Big) , \\
   \hat{E}^{ab} &= \ee^\Delta \left( e^{ab} + A\wedge e^{ab} 
      - j\tA\wedge e^{ab}
      + \tfrac{1}{2}jA\wedge A \wedge e^{ab} \right) , \\
   \hat{E}^{a_1\dots a_5} &= \ee^{\Delta} \left( e^{a_1\dots a_5} 
      + jA\wedge e^{a_1\dots a_5} \right) , \\
   \hat{E}^{a,a_1\dots a_7} &= \ee^\Delta e^{a,a_1\dots a_7} . 
\end{aligned}
\end{equation}
In this frame, the components of the generalised vector
\begin{equation}
   V = v^a \hat{E}_a + \tfrac{1}{2} \omega_{ab} \hat{E}^{ab} 
         + \tfrac{1}{5!}\sigma_{a_1\dots a_5} \hat{E}^{a_1\dots a_5} 
         + \tfrac{1}{7!}\tau_{a,a_1\dots a_7} \hat{E}^{a,a_1\dots a_7} 
\end{equation}
can be used to construct 
\begin{equation}
\label{eq:split-iso}
\begin{aligned}
   V^{(A,\tA,\Delta)}
      &= v^a \hat{e}_a + \tfrac{1}{2} \omega_{ab} e^{ab} 
         + \tfrac{1}{5!}\sigma_{a_1\dots a_5} e^{a_1\dots a_5} 
         + \tfrac{1}{7!}\tau_{a,a_1\dots a_7} e^{a,a_1\dots a_7} \\
       &\in\Gs{TM \oplus \Lambda^2T^*M \oplus \Lambda^5T^*M
         \oplus (T^*M \otimes\Lambda^7T^*M)} ,
\end{aligned}
\end{equation}
thus realising the isomorphism~\eqref{eq:Eiso}. 

Given the generalised structure bundle one can then define vector
bundles associated to any given representation of
$\Edd\times\bbR^+$. We refer to sections of such bundles as
\emph{generalised tensors}. 


\subsection{The Dorfman derivative}
\label{sec:gen-lie}

The generalised tangent space admits a generalisation of
the Lie derivative which ultimately will encode the local bosonic
symmetries of the supergravity. Given
$V=v+\omega+\sigma+\tau\in\Gs{E}$, one can define an operator 
$\Lgen_V$, the \emph{Dorfman derivative}, which combines the action of
an infinitesimal diffeomorphism generated by $v$ and $A$- and
$\tA$-field gauge transformations generated by $\omega$ and
$\sigma$. In components, acting on $V' \in \Gs{E}$, it is given by
\begin{equation}
\label{eq:Lgen}
\begin{aligned} 
   \Lgen_V V' &= \mathcal{L}_v v' 
       + \left( \mathcal{L}_v \omega' - i_{v'} \dd\omega \right)
       + \left( \mathcal{L}_v \sigma' - i_{v'} \dd\sigma
          - \omega'\wedge\dd\omega \right) 
       \\ & \qquad 
       + \left( \mathcal{L}_v \tau'
          - j\sigma'\wedge\dd\omega
          - j\omega'\wedge\dd\sigma \right) ,
\end{aligned}
\end{equation}
where $\mathcal{L}_v$ is the conventional Lie derivative. It can also
be written in an $\Edd\times\bbR^+$ form, using coordinate indices
$M$, as 
\begin{equation}
\label{eq:Lgen-cov}
   \Lgen_V V^{\prime M}
      = V^N \der_N V^{\prime M} 
         - (\der\oadj V)^M{}_N V^{\prime N} ,
\end{equation}
where the action of the partial derivative operator has been embedded
into the dual generalised tangent space via the map $T^*M\to E^*$ so
that 
\begin{equation}
\label{eq:d-def}
   \der_M 
      = \begin{cases} \der_m  & \text{for $M=m$} \\
         0 & \text{otherwise}
         \end{cases} , 
\end{equation}
and $\oadj$ is the projection to the adjoint representation of
$\Edd\times\bbR^+$ 
\begin{equation}
   \label{eq:oadj}
   \oadj : E^* \otimes E \to \adj{\tilde{F}} ,
\end{equation}
as defined in~\eqref{eq:EE*-adj}. By taking the appropriate adjoint
action on the given $\Edd\times\bbR^+$ representation, the Dorfman
derivative can be naturally extended to an arbitrary generalised
tensor.  


\subsection{Generalised $\Edd\times\bbR^+$ connections and torsion}
\label{sec:gen-con}

\emph{Generalised connections} are first-order linear differential
operators $\Dgen$, analogues of conventional connections on $TM$,
which can be written in the form, given $W=W^A\hat{E}_A\in\Gs{E}$ in
frame indices 
\begin{equation}
   \Dgen_M W^A = \der_M W^A + \Omega_M{}^A{}_B W^B , 
\end{equation}
where $\Omega$ is a section of $E^*$ (denoted by the $M$ index) taking
values in $\adj{\tilde{F}}$ (denoted by the $A$ and $B$ frame
indices), and as such, the action of $\Dgen$ then extends naturally to
any generalised $\Edd\times\bbR^+$ tensor. Note that unlike a
conventional connection, the index $M$ runs over the whole of $E^*$
and so one can take the derivative not only in a vector direction but
also along two-forms, five-forms and so on.  

Let $\alpha$ be a generalised tensor and $\Lgen^\Dgen_V\alpha$ be its
Dorfman derivative~\eqref{eq:Lgen-cov} with $\der$ replaced by
$\Dgen$. The \emph{generalised torsion} of the generalised connection
$D$ can be defined as a linear map $T:\Gs{E}\to\Gs{\adj(\tilde{F})}$
given by 
\begin{equation}
\label{eq:Tgen-def}
   T(V)\cdot \alpha 
       = \Lgen^\Dgen_V \alpha - \Lgen_V \alpha , 
\end{equation}
for any $V\in\Gs{E}$ and where $T(V)$ acts via the adjoint
representation on $\alpha$. Remarkably one finds that the torsion is
an element of only particular irreducible representations of
$E^*\otimes\adj{\tilde{F}}$ as listed in table~\ref{tab:gen-torsion}. 
\begin{table}[htb]
\begin{center}
\begin{tabular}{ll}
   $\Edd$ group & torsion rep. \\
   \hline
   $E_{7(7)}$ & $\rep{912}_\rep{-1} + \rep{56}_\rep{-1}$ \\
   $E_{6(6)}$ & $\rep{351}'_\rep{-1} + \rep{27}_\rep{-1}$ \\
   $E_{5(5)}\simeq\Spin(5,5)$  
     & $\rep{144}^c_\rep{-1} + \rep{16}^c_\rep{-1}$ \\
   $E_{4(4)}\simeq\SL(5,\bbR)$ 
     & $\rep{40}_\rep{-1} + \rep{15}'_\rep{-1} + \rep{10}_\rep{-1}$ 
\end{tabular}
\end{center}
\caption{Generalised torsion representations}
\label{tab:gen-torsion}
\end{table}
As discussed in~\cite{CSW2}, these are exactly the representations
that appear in the embedding tensor formulation of gauged
supergravities~\cite{embedT1}, including gaugings~\cite{embedT2} of
the so-called ``trombone'' symmetry~\cite{trombone}. 

We can construct a simple example of a generalised connection with
torsion as follows. Let $\nabla$ be a conventional connection. Given a
conformal split frame, it can be lifted to a generalised connection
acting on $E$ by taking 
\begin{equation}
\label{eq:Dnabla}
   \Dgen^\nabla_M V
    = \begin{cases}
         \begin{aligned}
            &(\nabla_m v^a) \hat{E}_a 
            + \tfrac{1}{2} (\nabla_m \omega_{ab}) \hat{E}^{ab} 
            \\ & \quad
            + \tfrac{1}{5!} (\nabla_m \sigma_{a_1\dots a_5}) 
               \hat{E}^{a_1\dots a_5} 
            + \tfrac{1}{7!} (\nabla_m \tau_{a,a_1\dots a_7})
               \hat{E}^{a,a_1\dots a_7} 
         \end{aligned}
      & \text{for $M=m$} \\
      0  & \text{otherwise} 
      \end{cases} . 
\end{equation}
By construction $\Dgen^\nabla$ depends on a choice of $A$, $\tA$ and
$\Delta$ used to define the frame as well of $\nabla$. The generalised
torsion of $\Dgen^\nabla$ is then given by 
\begin{equation}
\label{eq:splitT}
   T(V) = \ee^\Delta \big(- i_v \dd\Delta + v\otimes \dd\Delta
              - i_v F + \dd \Delta \wedge \omega 
              - i_v \tF + \omega \wedge F 
              + \dd \Delta \wedge \sigma \big) , 
\end{equation}
using the notation of~\eqref{eq:Ed-secs}. For other examples of
generalised connections, with and without torsion, see
also~\cite{GLSW,GO1}.


\subsection{Generalised $G$ structures}

In what follows we will be interested in further refinements of the
generalised frame bundle $\tilde{F}$. We define a \emph{generalised $G$
structure} $P$ as a $G\subset \Edd\times\bbR^+ $ principle sub-bundle of
the generalised structure bundle $\tilde{F}$, that is  
\begin{equation}
\label{eq:G-structure}
   P \subset \tilde{F} \text{ with fibre $G$}.
\end{equation}
It picks out a special subset of frames that are related by $G$
transformations. Typically one can also define $P$ by giving a set of
nowhere vanishing generalised tensors $\{K_{(a)}\}$, invariant under
the action of $G$. By definition, the invariant tensors parametrise,
at each point $x\in M$, an element of the coset 
\begin{equation}
   \left. \{K_{(a)}\} \right|_x \in \frac{\Edd \times\bbR^+}{G} . 
\end{equation}
A generalised connection $\Dgen$ is said to be compatible with the $G$
structure $P$ if it preserves all the invariant tensors  
\begin{equation}
   \Dgen K_{(a)} = 0
\end{equation}
or, equivalently, if the derivative acts only in the $G$ sub-bundle $P$.

A special class of generalised $G$ structures are those characterised
by the maximal compact subgroup $\Hd$ of $\Edd$. In the next section
we shall see how the extra data present in an $\Hd$ structure allows one
to naturally describe eleven-dimensional supergravity. 


\section{Supergravity as $\Hd$ generalised gravity}
\label{sec:GenGrav}

We now turn to the main result of this paper. We give a complete
rewriting in the language of generalised geometry of the restricted
eleven-dimensional supergravity, to leading order in fermions. This
will result in a unified formulation which has the larger bosonic
symmetries of the theory manifest. Specifically, the local symmetry of
the theory is $\Spin(10-d,1)\times\dHd$ where $\dHd$ is the
double-cover of the maximal compact subgroup of $\Edd$. 


\subsection{Supergravity degrees of freedom and $\Hd$ structures}
\label{sec:Hd}

\subsubsection{Bosons}

As discussed in~\cite{CSW2}, the bosonic supergravity fields define a
generalised $\Hd$ structure $P$, where $\Hd$ is the maximally compact
subgroup of $\Edd$. These, or rather their double
covers\footnote{Since the underlying manifold $M$ is assumed to
   possess a spin structure, we are free to promote to the double
   cover.} $\dHd$, are listed in table~\ref{tab:coset}.  
\begin{table}[htb]
\begin{center}
\begin{tabular}{llll}
   $\Edd$ group & $\dHd$ group & $E\simeq E^*$ 
      & $\adj{P}^\perp = \adj{\tilde{F}}/\adj{P}$ \\
   \hline
   $E_{7(7)}$ & $SU(8)$ & $\rep{28}+\bar{\rep{28}}$ 
      & $\rep{35}+\rep{\bar{35}}+\rep{1}$ \\
   $E_{6(6)}$ & $\textit{USp}(8)$ & $\rep{27}$
      & $\rep{42}+\rep{1}$ \\
   $E_{5(5)}\simeq\Spin(5,5)$  & $\Spin(5)\times\Spin(5)$ 
      & $\repp{4}{4}$
      & $\repp{5}{5}+\repp{1}{1}$ \\
   $E_{4(4)}\simeq\SL(5,\bbR)$ & $\Spin(5)$ & $\rep{10}$ 
      & $\rep{14}+\rep{1}$ 
\end{tabular}
\end{center}
\caption{Double covers of the maximal compact subgroups of $\Edd$ and
  $\Hd$ representations of the generalised tangent spaces and coset
  bundles}
\label{tab:coset}
\end{table}

An $\Hd$ structure on $\tilde{F}$ is the direct analogy of a metric
structure, where one considers the set of orthonormal frames related
by $O(d)$ transformations. The choice of such a structure is
parametrised, at each point on the manifold, by a Riemannian metric
$g$, a three-form $A$ and a six-form $\tA$ gauge fields, and a scalar
$\Delta$, that is
\begin{equation}
   \{ g, A, \tA, \Delta \} 
      \in \frac{\Edd \times\bbR^+}{\Hd} . 
\end{equation}
These are precisely the set of bosonic fields in the restricted
theory. The corresponding coset representations are listed in
table~\ref{tab:coset}. 

One can construct elements of the structure bundle $P\subset\tilde{F}$
concretely, that is, identify the analogues of ``orthonormal'' frames, 
as follows. It is always possible to choose an $\Hd$ frame in a
conformal split form~\eqref{eq:geom-basis}, where now one takes
$\hat{e}_a$ to be an orthonormal basis of $TM$ for the metric $g$. Any
other frame is then related by an $\Hd$ transformation. (The action of
$\Hd$ on the generalised tangent space is given explicitly
in~\eqref{eq:E-tranfs} and~\eqref{eq:Hd-embed}.) 

As in the Riemannian case, one can also also construct a
\emph{generalised metric}, which is invariant under a change of $\Hd$
frame. Given $V=V^A \hat{E}_A\in\Gs{E}$, expanded in an $\Hd$ basis,
one defines   
\begin{equation}
\label{eq:flat-gen-metric}
   G(V,V) = v^2 + \tfrac{1}{2!}\omega^2 
      + \tfrac{1}{5!}\sigma^2 + \tfrac{1}{7!}\tau^2 ,
\end{equation}
where $v^2=v_av^a$, $\omega^2=\omega_{ab}\omega^{ab}$,
$\sigma^2=\sigma_{a_1\dots a_5}\sigma^{a_1\dots a_5}$,
$\tau^2=\tau_{a,a_1\dots a_7} \tau^{a,a_1\dots a_7}$, and indices are
contracted using the flat frame metric $\delta_{ab}$. (Note that $G$
allows us to identify $E\simeq E^*$.)  It is easy to show, given the
transformation~\eqref{eq:E-tranfs}, that this is an $\Hd$ 
invariant, independent of the choice of $\Hd$ frame. Thus it can be
evaluated in the conformal split representative~\eqref{eq:geom-basis}
and one sees explicitly that the metric is defined by the fields $g$,
$A$, $\tA$ and $\Delta$ that determine the coset element.
Explicit expressions for the generalised metric in terms of the
supergravity fields in the coordinate frame have been worked out,
for example, in~\cite{BP}. The fact that there is always a singlet
present in the coset representations, as can be seen from
table~\ref{tab:coset}, implies that there is always a density that is
$\Hd$ (and $\Edd$) invariant, corresponding to the choice of $\bbR^+$
factor and which we denote as $\volG$. In a coordinate frame it is
given by\footnote{In general, $\volG$ can be related to the
   determinant of the metric by $\det{G}=\volG ^{-\dim{E}/(9-d)}$.}  
\begin{equation}
\label{eq:Phi-def}
   \volG = \sqrt{g}\, \ee^{(9-d)\Delta} . 
\end{equation}

As described in~\cite{CSW2, GMPW}, the infinitesimal bosonic symmetry
transformation is naturally encoded as the Dorfman derivative by $V
\in\Gs{E}$  
\begin{equation}
   \delta_V G = \Lgen_V G ,
\end{equation}
and the algebra of these transformations is given by
$\BLie{\Lgen_U}{\Lgen_V}=\Lgen_{\Lgen_U{V}}=-\Lgen_{\Lgen_VU}=\Lgen_{\Bgen{U}{V}}$
where the Courant bracket $\Bgen{U}{V}$ is the antisymmetrisation of the
Dorfman derivative. 


\subsubsection{Fermions}

The fermionic degrees of freedom form spinor representations of
$\dHd$, the double cover\footnote{Note that, as discussed in
   appendix~\ref{app:invol}, $\dHd$ can be defined abstractly for all
   $d\leq8$ as the subgroup of $\Cliff(d;\bbR)$ preserving a
   particular involution of the algebra.} 
of $\Hd$~\cite{deWN,nic,keur}. Let $S$ and $J$ denote the bundles
associated to the representations of $\dHd$ listed in
table~\ref{tab:SJ}. The fermion fields $\psi$, $\rho$ and the
supersymmetry parameter $\varepsilon$ of the restricted theory are 
sections 
\begin{equation}
\begin{aligned}
\label{eq:gravitino-rep}
  \psi &\in \Gs{J} , &&&&&
  \rho &\in \Gs{S} , &&&&&
  \varepsilon &\in \Gs{S} .
\end{aligned}
\end{equation}
\begin{table}[htb]
\begin{center}
\begin{tabular}{lll}
	$\dHd$ & $S$ & $J$ \\ 
	\hline
	$\SU(8)$ & $\rep{8}+\bar{\rep{8}}$ & $\rep{56}+\bar{\rep{56}}$ \\
	$U\! Sp(8)$ & $\rep{8}$ & $\rep{48}$\\
	$\Spin(5)\times \Spin(5)$ & $\repp{4}{1} + \repp{1}{4}$ 
            & $\repp{4}{5} + \repp{5}{4}$\\
	$\Spin(5)$ & $\rep{4}$ & $\rep{16}$	
\end{tabular}
\end{center}
\caption{Spinor and gravitino representations in each dimension
\label{tab:SJ}}
\end{table}

However, the restricted fermions also transform as spinors of the
flat $\bbR^{10-d,1}$ space. As discussed in section~\ref{sec:sugra},
the simplest formulation is to view them as eleven-dimensional spinors
and use the embedding
$\Spin(10-d,1)\times\dHd\subset\Cliff(10,1;\bbR)$ described in
appendix~\ref{app:Hd-cliff101}\footnote{The alternative is to decompose the
   eleven-dimensional spinors which necessarily leads to dimension
   dependent expressions, as can be seen from
   appendix~\ref{app:spin-decomp}. That approach is therefore better
   suited for the explicit constructions we will be examining in the
   next section. For now we maintain the discussion completely
   general.}. This will allow us to write expressions directly 
comparable to the ones in section~\ref{sec:sugra}. There is a
complication, in that there are actually two distinct ways of
realising the action of $\dHd$ on the $\Cliff(10,1;\bbR)$ spinor bundle
$\hat{S}$, related by a change of sign of the gamma matrices. Given
$\chi^\pm\in\Gs{\hat{S}}$ and $N\in\Gs{\adj P}$ we have the two actions
\begin{equation}
   N\cdot\hat{\chi}^\pm 
       = \tfrac{1}{2}\big(\tfrac{1}{2!}n_{ab}\Gamma^{ab} 
             \pm \tfrac{1}{3!}b_{abc}\Gamma^{abc} 
             - \tfrac{1}{6!}\tb_{a_1\dots a_6}\Gamma^{a_1\dots a_6}
          \big) \hat{\chi}^\pm . 
\end{equation}
If one denotes as $\hat{S}^\pm$ the bundle of spinors
transforming under the two actions, one finds, for even $d$, that the
two representations are equivalent, and
$\hat{S}\simeq\hat{S}^+\simeq\hat{S}^-$. However for odd $d$ they are
distinct and the spinor bundle decomposes
$\hat{S}\simeq\hat{S}^+\oplus\hat{S}^-$. The same applies to
spin-$\tfrac{3}{2}$ bundles $\hat{J}^\pm$. The
$\Spin(10-d,1)\times\dHd$ representations of the corresponding four 
bundles listed in table~\ref{tab:SJodd} (see also~\cite{gen-holo-duff}). 
\begin{table}[htb]
\begin{center}
\begin{tabular}{lcccc}
   $d$ & $\hat{S}^-$ & $\hat{S}^+$ & $\hat{J}^-$ & $\hat{J}^+$ \\
   \hline
   $7$ & $\repp{2}{8}+\repp{\bar{2}}{\bar{8}}$ 
      & $\repp{2}{\bar{8}}+\repp{\bar{2}}{8}$ 
      & $\repp{2}{56}+\repp{\bar{2}}{\bar{56}}$ 
      & $\repp{2}{\bar{56}}+\repp{\bar{2}}{56}$ \\
   $6$ & $\repp{4}{8}$ & $\repp{4}{8}$ & $\repp{4}{48}$ & $\repp{4}{48}$ \\
   $5$ & $(\rep{4},\rep{4},\rep{1}) + (\bar{\rep{4}},\rep{1},\rep{4}) $ 
      & $(\rep{4},\rep{1},\rep{4}) + (\bar{\rep{4}},\rep{4},\rep{1})$ 
      & $(\rep{4},\rep{4},\rep{5}) + (\bar{\rep{4}},\rep{5},\rep{4})$ 
      & $(\rep{4},\rep{5},\rep{4}) + (\bar{\rep{4}},\rep{4},\rep{5})$ \\
   $4$ & $\repp{8}{4}$ & $\repp{8}{4}$ 
      & $\repp{8}{16}$ & $\repp{8}{16}$ 
\end{tabular}
\end{center}
\caption{Spinor and gravitino as $\Spin(10-d,1)\times\dHd$ representations. Note that when $d$ is even the positive and negative representations are actually equivalent.
\label{tab:SJodd}}
\end{table}

Finally, we find that the supergravity fields of
section~\ref{sec:sugra} can be identified as follows, 
\begin{equation}
\label{eq:gen-fermi}
\begin{aligned}
   \hat{\varepsilon}^- &= \ee^{-\Delta/2} \;\varepsilon^{\text{sugra}}
      && \in\Gs{\hat{S}^-} ,\\
   \hat{\rho}^+ &= \ee^{\Delta/2} \;\rho^{\text{sugra}} 
      && \in \Gs{\hat{S}^+} ,\\
   \hat{\psi}_a^- &= \ee^{\Delta/2} \;\psi_a^{\text{sugra}}  \
      && \in \Gs{\hat{J}^-} .
\end{aligned}
\end{equation}
Note that, due to the warping of the metric, the precise maps between
the fermion fields as viewed in the geometry and in the supergravity
description involve a conformal rescaling. This is of course purely
conventional, since one could just as easily perform field
redefinitions at the supergravity level. We chose, however, to
maintain the conventions in section~\ref{sec:sugra} as familiar as
possible and make the identification at this point. 


\subsection{Supergravity operators}
\label{sec:genLC}

The differential operators and curvatures that appear in the
supergravity equations will be built out of generalised connections
$\Dgen$ which are simultaneously torsion-free and $\Hd$ compatible, in
analogy to the Levi--Civita connection. Recall that a generalised
connection is said to be compatible with the $\Hd$ structure if $\Dgen
G = 0$ or, equivalently, if the derivative acts only in the $\Hd$
sub-bundle. We proved in~\cite{CSW2} that there always exists such a
torsion-free, metric compatible connection but, unlike the
Levi--Civita connection, it is not unique. 
     
To see this, let $\nabla$ be the Levi--Civita connection for the
metric $g$ and $\Dgen^\nabla$ its lift to an action on $E$ as
in~\eqref{eq:Dnabla}. Since $\nabla$ is compatible with the
$O(d)\subset\Hd$ subgroup, it is necessarily an $\Hd$-compatible
connection. However, $\Dgen^\nabla$ is not torsion-free, as can be
seen from~\eqref{eq:splitT}. To construct a torsion-free compatible
connection one simply modifies $\Dgen^\nabla$. A generic generalised
connection $\Dgen$ can always be written as  
\begin{equation}
\label{eq:metric+torsionfree}
   \Dgen_M W^ A = \Dgen^\nabla_M W^A + \Sigma_M{}^A{}_B W^B . 
\end{equation}
If $\Dgen$ is compatible with the $\Hd$ structure then
\begin{equation}
   \Sigma \in \Gs{E^*\otimes\adj{P}} ,
\end{equation}
that is, it is a generalised covector taking values in the adjoint of
$\Hd$. In~\cite{CSW2} we showed that one can always find a suitable $\Sigma$ such that the torsion of $\Dgen$ vanishes, but the solution is
not unique. Contracting with $V\in\Gs{E}$ so $\Sigma(V)\in\Gs{\adj{P}}$ and
using the basis for the adjoint of $\Hd$ given
in~\eqref{eq:Hd-alg} and~\eqref{eq:Hd-embed}, one finds that in a conformal split frame
\begin{equation}
\label{eq:gen-D}
\begin{aligned}
   \Sigma(V)_{ab} &= \ee^\Delta \left( 
        \tfrac{2(7-d)}{d-1}  v_{[a} \der_{b]} \Delta
        + \tfrac{1}{4!} \omega_{cd} F^{cd}{}_{ab} 
        + \tfrac{1}{7!} \sigma_{c_1 \dots c_5} 
           \tilde{F}^{c_1 \dots c_5}{}_{ab} 
        + \am(V)_{ab} \right) ,\\
   \Sigma(V)_{abc} &= \ee^\Delta \left( 
        \tfrac{6}{(d-1)(d-2)} (\dd \Delta \wedge \omega)_{abc} 
   	+ \tfrac14 v^d F_{dabc} + \am(V)_{abc} \right) ,\\
   \Sigma(V)_{a_1\dots a_6} &= \ee^\Delta \left( 
        \tfrac{1}{7} v^b \tilde{F}_{ba_1\dots a_6} 
        + \am(V)_{a_1\dots a_6} \right) , 
\end{aligned}
\end{equation}
where $\am\in\Gs{E^*\otimes\adj{P}}$ is the undetermined part of the
connection -- it projects to zero under the map to the torsion
representations. Clearly, requiring metricity and vanishing torsion is
not enough to specify a single generalised connection. 

Although $\Dgen$ is ambiguous, one can define projections of $\Dgen$
which result in unique operators. We identified four such maps
in~\cite{CSW2}, and they turn out to be directly related to the
representations of the fermion fields.  Since we are interested in 
comparing with the supergravity expressions, we can take the
embedding~\eqref{eq:Hd-cliff101} and consider the natural action of
$\Dgen$ on the $\Spin(10-d,1)\times\dHd$ representations listed in
table~\ref{tab:SJodd}. Following the notation
of~\eqref{eq:SOd-unique-ops} we define the projected operators
\begin{equation}
\label{eq:unique-ops}
\begin{aligned}
   \DSS {}&: \hat{S}^\pm \to \hat{S}^\mp , & && &&
   \DJJ {}&: \hat{J}^\pm \to \hat{J}^\mp , \\
   \DSJ {}&: \hat{S}^\pm \to \hat{J}^\pm , & && &&
   \DJS {}&: \hat{J}^\pm \to \hat{S}^\pm . 
\end{aligned}
\end{equation}
We can check that they are indeed independent of $\am$ by decomposing
under $\Spin(d)\subset\dHd$ and taking the torsion-free
connection~\eqref{eq:gen-D}. Using the formulae for the projections 
given in~\eqref{eq:SO-1st-proj} and~\eqref{eq:SO-2nd-proj}, and
already applying the operators to the supersymmetry parameter
$\hat{\varepsilon}^-$ in~\eqref{eq:gen-fermi}, we then find 
\begin{equation}
\begin{aligned}
   \DSS \hat{\varepsilon}^- 
      &= \ee^{\Delta/2} \Big( \slashed{\LC} 
         + \tfrac{9-d}{2} (\slashed{\der} \Delta) 
         - \tfrac{1}{4} \slashed{F} 
         - \tfrac{1}{4}  \slashed{\tF}  
         \Big) \varepsilon^{\text{sugra}}, \\
   (\DSJ \hat{\varepsilon}^-)_a 
      &= \ee^{\Delta/2} \Big( \LC_a 
         + \tfrac{1}{288}  (\Gamma_a{}^{b_1 \dots b_4} 
         - 8 \delta_a{}^{b_1} \Gamma^{b_2 b_3 b_4} ) F_{b_1 \dots b_4} 
      \\ & \hspace*{14em}
        - \tfrac{1}{12} \tfrac{1}{6!}  \tF_{ab_1 \dots b_6} 
            \Gamma^{b_1 \dots b_6}   
         \Big)\varepsilon^{\text{sugra}} .
\end{aligned}
\end{equation}
From derivatives of elements $\Gs{\hat{J}^\pm}$ we obtain the second
set of unique operators which using~\eqref{eq:SO-3rd-proj}
and~\eqref{eq:SO-4th-proj} as applied to $\hat{\psi}^-$
of~\eqref{eq:gen-fermi}, take the form
\begin{equation}
\begin{aligned}
   \DJS \hat{\psi}^-  
      &= \ee^{3\Delta/2}\Big[
         \LC^a - \tfrac{1}{10-d} \Gamma^{ab} \LC_b
         + (10-d) \der^a\Delta - \Gamma^{ab}\der_b \Delta
      \\ & \qquad \qquad
         - \tfrac{1}{4} \tfrac{1}{3!} F^a{}_{b_1 b_2 b_3} 
            \Gamma^{b_1 b_2 b_3} 
         + \tfrac{1}{4} \tfrac{1}{10-d} \tfrac{1}{4!} 
            \Gamma^a{}_{b_1 \dots b_4} F^{b_1 \dots b_4} 
      \\ & \qquad \qquad
         -\tfrac{1}{4} \tfrac{1}{6!} \tilde{F}^a{}_{b_1 \dots b_6} 
            \Gamma^{b_1 \dots b_6}
            \Big] \psi_a^{\text{sugra}} , \\
    (\DJJ \hat{\psi}^-)_a  
       &= -\ee^{3\Delta/2} \Big[ 
          \Gamma^c(\LC_c + \tfrac{11-d}{2} \der_c \Delta) \delta_a{}^b 
          + \tfrac{2}{9-d} \Gamma^b(
             \LC_a + \tfrac{11-d}{2} \der_a \Delta)
      \\ & \qquad \qquad
          - \tfrac{1}{12} (3+\tfrac{2}{9-d}) \slashed{F} \delta_a{}^b 
          + \tfrac13 \tfrac{10-d}{9-d} \tfrac{1}{2!} 
             F_{a}{}^b{}_{cd}\Gamma^{cd}  
      \\ & \qquad \qquad
          - \tfrac13 \tfrac{1}{9-d} \tfrac{1}{3!} 
             F_a{}^{c_1 \dots c_3} \Gamma^b{}_{c_1 \dots c_3} 
          + \tfrac16 \tfrac{10-d}{9-d} \tfrac{1}{3!} 
             F^{bc_1 \dots c_3} \Gamma_{a c_1\dots c_3} 
      \\ & \qquad \qquad
          - \tfrac{1}{6} \tfrac{1}{9-d} \tfrac{1}{4!}
             F_{c_1 \dots c_4}  \Gamma_a{}^{bc_1 \dots c_4}
          + \tfrac{1}{4} \tfrac{1}{5!} 
             \tF_{a}{}^b{}_{c_1 \dots c_5} \Gamma^{c_1 \dots c_5}  
             \Big] \psi_b^{\text{sugra}}  .
\end{aligned}
\end{equation}
These four operators, all constructed from the same connection, will
now enable us to rewrite all the supergravity equations of
section~\ref{sec:eom}. 


\subsection{Supergravity equations from generalised geometry}
\label{sec:abstract-sugra}


\subsubsection{Supersymmetry algebra}

Comparing with~\eqref{eq:susy-ferm}, we immediately see that the
operators~\eqref{eq:unique-ops} give precisely the supersymmetry
variations of the two fermion fields 
\begin{equation}
\label{eq:gen-susy-fermions}
\begin{aligned}
   \delta \hat{\psi}^- &= \DSJ \hat{\varepsilon}^-  , \\
   \delta \hat{\rho}^+ &= \DSS \hat{\varepsilon}^-  .
\end{aligned}
\end{equation}

Since the bosons arrange themselves into the generalised metric, one
expects that their supersymmetry variations~\eqref{eq:susy-bos} are
given by the variation of $G$. In fact, the most convenient object to
consider is $G^{-1} \delta G$ which is naturally a section of the
bundle $\adj(P)^\perp$, listed in table~\ref{tab:coset}. One has the
isomorphism~\eqref{eq:H-perp}
\begin{equation}
   \adj(P)^\perp \simeq 
      \bbR \oplus S^2 T^*M \oplus \Lambda^3 T^*M 
      \oplus \Lambda^6 T^*M
\end{equation}
and we can identify the component variations of the generalised metric,
as written in the split frame, as 
\begin{equation}
\begin{aligned}
   (G^{-1} \delta G) &= -2 \delta \Delta , \\
   (G^{-1} \delta G)_{ab} &= \delta g_{ab} \\
   (G^{-1} \delta G)_{abc} &= - \delta A_{abc} , \\
   (G^{-1} \delta G)_{a_1 \dots a_6} &= -\delta \tilde{A}_{a_1 \dots a_6} .
\end{aligned}
\end{equation}
One finds that the supersymmetry variations of the
bosons~\eqref{eq:susy-bos} can be written in the $\dHd$ covariant form  
\begin{equation}
\label{eq:gen-susy-bosons}
   G^{-1} \delta G 
      = (\hat\psi^-\!\proj{\,\adj{P}^\perp}\!\hat\varepsilon^- )
         + (\hat\rho^+\!\proj{\,\adj{P}^\perp}\!\hat\varepsilon^- ) ,
\end{equation}
where $\proj{\,\adj{P}^\perp}$ denotes the projection to
$\adj(P)^\perp$ given in~\eqref{eq:SJ-perp-proj}
and~\eqref{eq:SS-perp-proj}.


\subsubsection{Generalised Curvatures and the Equations of Motion}

To realise the fermionic equations of motion one uses the unique
projections~\eqref{eq:unique-ops}. We can then formulate the two
equations~\eqref{eq:eom-psi} and~\eqref{eq:eom-rho} as, respectively,  
\begin{equation}
\label{eq:gen-fermions}
\begin{aligned}
   -\DJJ \hat{\psi}^- 
      - \tfrac{11-d}{9-d} \DSJ \hat{\rho}^+ &= 0, \\
   -\DSS \hat{\rho}^+ - \DJS \hat{\psi}^- &= 0. 
\end{aligned}
\end{equation}
Note that $\hat\rho^+$ is embedded with a different conformal factor
to $\hat\varepsilon^-$ and also is a section of $\hat{S}^+$ rather
than $\hat{S}^-$. This means we have 
\begin{equation}
\begin{aligned}
   \DSS \hat\rho^+
     &= -\ee^{3\Delta/2} \Big( \slashed{\LC} 
        + \tfrac{11-d}{2} (\slashed{\der} \Delta) 
        + \tfrac{1}{4} \slashed{F} 
        - \tfrac{1}{4}  \slashed{\tF}  
        \Big)\rho^{\text{sugra}} \\
   (\DSJ \hat\rho^+)_a 
      &= \ee^{3\Delta/2} \Big[ 
         (\LC_a + \der_a \Delta)
         -\tfrac{1}{288}  (\Gamma_a{}^{b_1 \dots b_4} 
         - 8 \delta_a{}^{b_1} \Gamma^{b_2 b_3 b_4} ) F_{b_1 \dots b_4} 
         \\ & \hspace*{12em}
         - \tfrac{1}{12} \tfrac{1}{6!}
            \tF_{ab_1 \dots b_6} \Gamma^{b_1 \dots b_6} 
         \Big] \rho^{\text{sugra}}
\end{aligned}
\end{equation}

From these we can find the generalised Ricci tensor $\GenRicci_{AB}$,
following~\cite{CSW2}. Recall that the supersymmetric variation of the
fermionic equations of motion vanishes up to the bosonic equations of
motion~\eqref{eq:gen-fermions}.  Anticipating that the bosonic
equations of motion will correspond to $\GenRicci_{AB}=0$, one way to
define generalised Ricci tensor is via the variation
of~\eqref{eq:gen-fermions} under~\eqref{eq:gen-susy-fermions}. By
construction this gives  $\GenRicci_{AB}$ as a section of
$\adj{P}^\perp\subset E^*\otimes E^*$, the same space as variations of
the generalised metric $\delta G$, in complete analogy to the
conventional metric and Ricci tensor. Defining $\GenRicci_{AB}$ as an
$\dHd$ tensor we write
\begin{equation}
\label{eq:GenRS}
\begin{aligned}
   -\DJJ (\DSJ \hat{\varepsilon}^-) 
      -\tfrac{11-d}{9-d} \DSJ 
         (\DSS \hat{\varepsilon}^-) 
      &= \GenRic \cdot \hat{\varepsilon}^-, \\
   \DJS (\DSJ \hat{\varepsilon}^-) 
      + \DSS (\DSS \hat{\varepsilon}^-) 
      &= \GenS\, \hat{\varepsilon}^-, \\[4pt]
\end{aligned}
\end{equation}
for any $\hat{\varepsilon}^- \in \Gs{\hat{S}^-}$ and where $\GenS$ and
$\GenRic_{AB}$ are the scalar and non-scalar parts of $\GenRicci_{AB}$
respectively. The action of $\GenRic_{AB}$ on
$\hat{\varepsilon}^-$ that appears of the right-hand side
of~\eqref{eq:GenRS} is given explicitly in~\eqref{eq:perp-S-proj}. 

In components, using the notation of~\eqref{eq:H-perp}, we find 
\begin{equation}
\begin{aligned}
   \GenS &= \ee^{2\Delta} \Big[ 
       \Scalar - 2(\dex-1) \LC^2 \Delta - \dex(\dex-1) (\der \Delta)^2 
       - \tfrac12 \tfrac{1}{4!} F^2 
       - \tfrac12 \tfrac{1}{7!} \tilde{F}^2 \Big] \\
   \GenRicci_{ab} &= \ee^{2\Delta} \Big[ 
       \Ric_{ab} - \dex \LC_a \LC_b \Delta 
       - \dex(\der_a \Delta)(\der_b \Delta) 
       \\ & \hspace{40pt} 
       - \tfrac12 \tfrac{1}{4!}\Big( 
          4F_{a c_1 c_2 c_3} F_b{}^{c_1 c_2 c_3}
          - \tfrac{1}{3} g_{ab} F^2 \Big) 
       \\ &\hspace{40pt}
       - \tfrac12 \tfrac{1}{7!}\left(  7\tilde{F}_{a c_1 \dots c_6} 
          \tilde{F}_b{}^{c_1 \dots c_6}
          - \tfrac{2}{3} g_{ab} \tilde{F}^2 \right) \Big] ,\\
    \GenRicci_{abc} &= \tfrac12 \ee^{2\Delta}*\Big[ 
       \ee^{-\dex \Delta} \dd * (\ee^{\dex \Delta}F) 
       -  F\wedge *\tilde{F} \Big]_{abc} ,\\
    \GenRicci_{a_1 \dots a_6} &= \tfrac12 \ee^{2\Delta} *\Big[
       \ee^{-\dex \Delta} \dd*(\ee^{\dex \Delta} \tilde{F}) 
       \Big]_{a_1 \dots a_6} ,
\end{aligned}
\end{equation}
where $\dex = 11-d$. The generalised Ricci tensor is manifestly
uniquely determined and comparing with~\eqref{eq:eom11d} we see that
the bosonic equations of motion become simply 
\begin{equation}
\label{eq:gen-bosons}
   \GenRicci_{AB} = 0 .
\end{equation}

The bosonic action~\eqref{eq:Boson-action} is given by the generalised curvature scalar, integrated with the volume form~\eqref{eq:Phi-def}
\begin{equation}
\label{eq:S-gen}
   S_{\text{B}} =\frac{1}{2\kappa^2} \int \volG \GenS .
\end{equation}
Finally, the fermionic action can be written using the natural invariant pairings of the terms in~\eqref{eq:gen-fermions} with the fermionic fields. Using the expressions~\eqref{eq:Hd-JJ-singlet} and~\eqref{eq:Hd-SS-singlet} for the spinor bilinears, we find that~\eqref{eq:Fermion-action} can be rewritten as
\begin{equation}
\begin{aligned}
\label{eq:gen-fermion-action}
S_{\text{F}} &= \frac{1}{\kappa^2} \int \volG \Big[
   -\bl{\hat{\psi}^-}{\DJJ\hat{\psi}^-}
   -\tfrac{c}{c-2} 
       \bl{\hat{\psi}^-}{\DSJ\hat{\rho}^+}
   \\ & \hspace*{9em}
   +\tfrac{c(c-1)}{(c-2)^2} 
       \bl{\hat{\rho}^+}{\DJS\hat{\psi}^-}
   + \tfrac{c(c-1)}{(c-2)^2}
       \bl{\hat{\rho}^+}{\DSS\hat{\rho}^+}
   \Big].
\end{aligned}
\end{equation}
%


\section{Explicit $\Hd$ constructions}
\label{sec:explicit}

 
In the previous section, we gave the generic construction of the
supergravity in terms of generalised geometry, valid in all  
$d\leq 7$. The theory has a local $\dHd$ symmetry, however this might not be
not totally explicit since we used a $\Cliff(10,1;\bbR)$ formulation for the
fermionic fields. 

For completeness, we now demonstrate for two examples, in $d=4$ and $d=7$, how one can write our expressions with indices which transform directly under the manifest local $\Spin(5)$ and $\SU(8)$ symmetries of the previous section. Correspondingly, in this section we treat the fermions slightly differently from the previous ones. Whereas before we kept all spinors as $\Cliff(10,1;\bbR)$ objects, we now want to make their $\dHd$ nature more explicit. In order to make this possible, one has to decompose the eleven-dimensional spinors following the procedures outlined in appendix~\ref{app:spin-decomp} and embed the $\Cliff(d;\bbR)$ expressions into $\dHd$ representations, according to appendix~\ref{app:Hd-Od}. We will then keep the external spinor indices of the fermion fields hidden and treat them as sections of the genuine $\dHd$ bundles $S$ and $J$.


\subsection{$d=4$ and $\tilde{H}_4=\Spin(5)$}
\label{sec:4d}


\subsubsection{$\GL^+(5,\bbR)$ generalised geometry}

In four dimensions, we have 
$E_{4(4)}\times\bbR^+\simeq\SL(5,\bbR)\times\bbR^+\simeq\GL^+(5,\bbR)$. We
can then write the generalised geometry explicitly in terms of indices
$i,j,k,\dots = 1, \dots, 5$ transforming under $\GL^+(5,\bbR)$. 
 
Generalised vectors $V$ transform in the antisymmetric $\rep{10}$
representation. We can introduce a basis $\{\hat{E}_{ii'}\}$ (locally a
section of the generalised structure bundle $\tilde{F}$) transforming
under $\GL^+(5,\bbR)$ so that 
\begin{equation}
   V = \tfrac{1}{2} V^{ii'} \hat{E}_{ii'} . 
\end{equation}
In the conformal split frame~\eqref{eq:geom-basis}, 
we can identify~\cite{chris,BP-alg}
\begin{equation}
\begin{aligned}
   \hat{E}_{a5} &= \ee^\Delta \left(
        \hat{e}_a + i_{\hat{e}_a} A \right) ,\\
   \hat{E}_{ab}  &= \tfrac{1}{2} \ee^\Delta \epsilon_{abcd} e^{cd} ,
\end{aligned}
\end{equation}
where $\epsilon$ is the numerical totally antisymmetric
symbol. Equivalently 
\begin{equation}
\begin{aligned}
   V^{a5} &= v^a ,\\
   V^{ab} &= \tfrac{1}{2} \epsilon^{abcd} \omega_{cd} ,
\end{aligned}
\end{equation}
where $v^a$ and $\omega_{ab}$ are as in~\eqref{eq:split-iso}. In this
frame the partial derivative~\eqref{eq:d-def} $\der_{ii'}$ has the
form 
\begin{equation}
\begin{aligned}
   \der_{a5} &= \tfrac12 \ee^\Delta \der_a  , \\
   \der_{ab} &= 0 . 
\end{aligned}
\end{equation}
Note that there is also a generalised tensor bundle $W$ which
transforms in the fundamental $\rep{5}$ representation of
$\GL^+(5,\bbR)$. One finds 
\begin{equation}
   W \simeq (\det T^*M)^{1/2}\otimes \left( 
       TM \oplus \det TM\right) , 
\end{equation}
and a choice of basis $\{\hat{E}_{ii'}$\} defines a basis
$\{\hat{E}_i\}$ of $W$ where $K=K^i\hat{E}_i \in\Gs{W}$, such that
\begin{equation}
   \hat{E}_{ii'} = \hat{E}_i\wedge\hat{E}_{i'} , 
\end{equation}
since $E\simeq\Lambda^2W$, and where we use the four-dimensional
isomorphism $\det{T^*M}\otimes\Lambda^2TM\simeq\Lambda^2T^*M$.  

With this notation we can then use the $\GL^+(5,\bbR)$ adjoint action
explicitly to write the Dorfman derivative~\eqref{eq:Lgen-cov} of a
generalised vector. It takes its simplest form in the coordinate
frame~\eqref{eq:coord}, where it reads 
\begin{equation}
\label{eq:4d-dorfman}
   \Lgen_V W^{ij}
      = V^{kk'} \der_{kk'} W^{ij}
        + 4 \left(\der_{kk'} V^{k[i}\right) W^{j]k'} 
      + \big(\der_{kk'} V^{kk'} \big) W^{ij} .
\end{equation}
This form of the $d=4$ Dorfman derivative was given, without the
$\bbR^+$ action, in~\cite{BP-alg}\footnote{For the 
   antisymmetrisation of $\Lgen_VW$ (which is simply the Courant bracket for
   two-forms~\cite{GCY}) in $\SL(5,\bbR)$ indices see also~\cite{sara}.}.
We can then write a generic generalised connection as 
\begin{equation}
\label{eq:4d-Dgen}
   \Dgen_{ii'} V^{jj'} = \partial_{ii'} V^{jj'} 
        + \Omega_{ii'}{}^{j}{}_k V^{kj'} 
        + \Omega_{ii'}{}^{j'}{}_k V^{jk} ,
\end{equation}
where the $j$ and $k$ indices of $\Omega_{ii'}{}^j{}_k$
parametrise an element of the adjoint of $\GL^+(5,\bbR)$.  


\subsubsection{$\Spin(5)$ structures and supergravity}
\label{sec:4d-expr}

In four dimensions $\Hd\simeq\SO(5)$ and we define the sub-bundle
$P\subset\tilde{F}$ of $\SO(5)$ frames as the set of frames where the
generalised metric~\eqref{eq:flat-gen-metric} can be written as
\begin{equation}
\label{eq:4d-gen-metric}
   G(V, W) = \tfrac{1}{2} \delta_{ij} \delta_{i'j'} V^{ii'} W^{jj'} ,
\end{equation}
where $\delta_{ij}$ is the flat $\SO(5)$ metric with which we can
raise and lower indices frame indices. Equivalently we can think of
the generalised metric as defining orthonormal frames on
the $\rep{5}$-representation bundle $W$.

Upon decomposing the fermionic fields of the supergravity according
to~\ref{app:spin-decomp-4d}, one finds that they embed into the spinor
and traceless vector-spinor representations of $\Spin(5)$. Our
conventions regarding $\Cliff(4;\bbR)$ and $\Cliff(5;\bbR)$ algebras
are given in appendix~\ref{app:cliff4} and we leave $\Spin(5)$ spinor
indices implicit throughout. We define
\begin{equation}
\label{eq:4d-fermion-def}
\begin{aligned}
   \varepsilon &= \ee^{- \Delta/2} 
      \varepsilon^{\text{sugra}}  &\in \Gs{S},\\
   \rho &= \ee^{\Delta/2} \gamma^{(4)} 
      \rho^{\text{sugra}} &\in \Gs{S},\\
   \psi_i &= 
      \begin{cases} 
         \ee^{\Delta/2} \gamma^{(4)}\left( 
            \delta^b{}_a 
            - \tfrac{2}{5} \gamma_a \gamma^b 
            \right)\psi_b^{\text{sugra}}   
            & \text{for } i=a \\
         - \tfrac{3}{5} \ee^{\Delta/2} 
            \gamma^a \psi_a^{\text{sugra}} 
            & \text{for } i = 5 
         \end{cases} 
         & \in \Gs{J}.\\
\end{aligned}
\end{equation}
Crucially, note the appearance of conformal factors in the
definitions, in similar fashion to~\eqref{eq:gen-fermi}. Recall also
that in four dimensions we have $S\simeq S^+\simeq S^-$, where the
action by $\gamma^{(4)}$ in the second line
of~\eqref{eq:4d-fermion-def} realises the second isomorphism. 

A generalised connection is compatible with the generalised
metric~\eqref{eq:4d-gen-metric} if $\Dgen G=0$. In terms of the
connection~\eqref{eq:4d-Dgen} in frame indices this implies
\begin{equation}
   \Omega_{ii'\,jj'} = - \Omega_{ii'\,j'j} 
\end{equation}
where indices are lowered using the $\SO(5)$ metric
$\delta_{ij}$. For such $\SO(5)$-connections, we can define the
generalised spinor derivative, given $\chi\in\Gs{S}$
\begin{equation}
   \Dgen_{ii'} \chi = \left(
        \der_{ii'} + \tfrac14 \Omega_{ii'jj'} \gamh^{jj'} \right) \chi .
\end{equation}
An example of such a generalised connection is the one~\eqref{eq:Dnabla}
defined by the Levi--Civita connection $\LC$, where, acting on
$\chi\in\Gs{S}$, we have 
\begin{equation}
   \Dgen^\LC_{ii'} \chi = \begin{cases}
         \tfrac{1}{2} \ee^\Delta \left(
            \der_{a} + \tfrac14 \omega_{a\,bc} \gamh^{bc} \right) \chi 
         & \text{if $i=a$ and $i'=5$} \\
         0 & \text{if $i=a$ and $i'=b$} 
         \end{cases} , 
\end{equation}
where $\omega_{a\,bc}$ is the usual spin-connection. 

We can construct a torsion-free compatible connection $\Dgen$, by
shifting $\Dgen^\LC$ by an additional connection piece
$\Sigma_{[ii'][jj']}$, such that its action on $\chi\in\Gs{S}$ is
given by 
\begin{equation}
   \Dgen_{ii'} \chi = \Dgen^\nabla_{ii'} \chi 
      + \tfrac14 \Sigma_{ii'jj'} \gamh^{jj'} \chi .
\end{equation}
The connection is torsion-free if 
\begin{equation}
   \Sigma_{ii'jj'} = \tfrac{1}{2} \left( 
           \delta_{j[i} \Sigma_{i']j'} - \delta_{j'[i} \Sigma_{i']j} \right) 
        + \am_{ii'jj'} ,
\end{equation}
where $\am_{ii'jj'}$ is the undetermined part -- traceless and symmetric under exchange of pairs of indices, so it transforms in the $\rep{35}$ of $\SO(5)$, see~\cite{CSW2} -- and
\begin{equation}
\begin{aligned}
   \Sigma_{a5} &= - \Sigma_{5a} = -2 \ee^\Delta \der_a \Delta ,\\
   \Sigma_{ab} &= - \tfrac{1}{12} \ee^\Delta F \delta_{ab} ,\\
   \Sigma_{55} &= \tfrac{7}{12} \ee^\Delta F ,
\end{aligned}
\end{equation}
with $F = \tfrac{1}{4!} \epsilon^{abcd} F_{abcd}$. The
projections~\eqref{eq:unique-ops} can be written in $\Spin(5)$ indices
as 
\begin{equation}
\label{4d-proj}
\begin{aligned}
   \DSS \varepsilon &= -\gamh^{ij} \Dgen_{ij} \varepsilon ,\\
   (\DSJ \varepsilon)_i &= 2 (\gamh^j \Dgen_{ij} \varepsilon
      - \tfrac15 \gamh_i \gamh^{jj'} \Dgen_{jj'} \varepsilon ) , \\
   (\DJJ \psi)_i &= - \gamh^{jj'} \Dgen_{jj'} \psi_i 
   	+\tfrac{12}{5}\Dgen_{ij} \psi^j  
	-\tfrac{8}{5} \gamh_{i}{}^{j} \Dgen_{jj'} \psi^{j'} ,\\
   \DJS \psi &= -\tfrac{5}{3}\gamh^{i} \Dgen_{ij} \psi^j .
\end{aligned}
\end{equation}
and are unique, independent of $\am_{ii'jj'}$. 

The supersymmetry variations of the
fermions~\eqref{eq:gen-susy-fermions} can then be written in a
manifestly $\Spin(5)$ covariant form  
\begin{equation}
\begin{aligned}
\label{eq:4dsusy-fermions}
   \delta \psi_i 
      &= (\DSJ \varepsilon)_i  
      = 2(\gamh^j \Dgen_{ij} \varepsilon 
         - \tfrac15 \gamh_i \gamh^{jj'} \Dgen_{jj'} \varepsilon ), \\
   \delta \rho 
      &= \DSS \varepsilon 
      = -\gamh^{ij} \Dgen_{ij} \varepsilon ,
\end{aligned}
\end{equation}
whereas the variation of the bosons~\eqref{eq:gen-susy-bosons} is
given by
\begin{equation}
   \delta G_{[ii'][jj']} 
      = \tfrac{1}{2} \big(
         \delta H_{i[j} \delta_{j']i'} 
         - \delta H_{i'[j} \delta_{j']i}  \big) , 
\end{equation}
with
\begin{equation}
   \delta H_{ij} = -2 \bar\varepsilon \gamh_{(i}  \psi_{j)}
      - \tfrac15 \delta_{ij} \bar\varepsilon \rho .
\end{equation}

Turning to the equations of motion, from~\eqref{eq:gen-fermions}, we
find that the fermionic equations take the form 
\begin{equation}
\begin{aligned}
\label{eq:4deom-fermions}
   -\tfrac{14}{5} (\gamh^j \Dgen_{ij} \rho 
      - \tfrac15 \gamh_i \gamh^{jj'} \Dgen_{jj'} \rho )
      - \gamh^{jj'} \Dgen_{jj'} \psi_i 
   	+\tfrac{12}{5}\Dgen_{ij} \psi^j  
	-\tfrac{8}{5} \gamh_{i}{}^{j} \Dgen_{jj'} \psi^{j'}   &= 0 ,\\
   \gamh^{ij} \Dgen_{ij} \rho + \tfrac{5}{3}\gamh^{i} \Dgen_{ij}
   \psi^j 
      &= 0 .
\end{aligned}
\end{equation}
The generalised Ricci tensor~\eqref{eq:GenRS}, after some
rearrangement and gamma matrix algebra, can be written as 
\begin{equation}
\begin{aligned}
   \GenRic_{ij} \gamh^j \varepsilon 
       &= \tfrac45 \gamh^j \BLie{\Dgen_{ik}}{\Dgen_j{}^k} \varepsilon
          - 2\gamh^{jkl} \BLie{\Dgen_{ij}}{\Dgen_{kl}}
             \varepsilon 
          - \tfrac{56}{25} \gamh_i{}^{jk}\BLie{\Dgen_{jl}}{\Dgen_k{}^l}
             \varepsilon 
       \\ & \qquad \qquad 
          - \tfrac{16}{5} \gamh^{jkl} \Dgen_{[ij} \Dgen_{kl]} \varepsilon
          + \tfrac85 \gamh_i{}^{j_1 \dots j_4} 
             \Dgen_{[j_1 j_2} \Dgen_{j_3j_4]} \varepsilon , \\
   \tfrac{5}{24} \GenS \varepsilon  
       &= \tfrac53 \gamh^{ii'jj'} \Dgen_{ii'} \Dgen_{jj'} \varepsilon 
	   - \tfrac53 \gamh^{ij} [ \Dgen_{ik} , \Dgen_j{}^k ] \varepsilon .
\end{aligned}
\end{equation}
Note that in this form one can clearly see that the curvatures cannot
be obtained simply from the commutator of two generalised covariant
derivatives. Instead, one must consider additional terms resulting
from a specific symmetric projection of the connections, as observed
in section~\textbf{3.3} of~\cite{CSW2}.  

The bosonic action~\eqref{eq:S-gen} is
\begin{equation}
   S_{\text{B}} =  \frac{1}{2\kappa^2} \int \volG \GenS .
\end{equation}
While the fermionic action~\eqref{eq:gen-fermion-action} can be
written as
\begin{equation}
\begin{aligned}
   S_{\text{F}} = \frac{1}{\kappa^2} \int \volG \Big( 
      &- \bar\psi^i ( - \gamh^{jk} \Dgen_{jk} \psi_i 
         + \tfrac{12}{5} \Dgen_{ij} \psi^j 
         - \tfrac85 \gamh_i{}^j \Dgen_{jk} \psi^k ) \\
      &- \tfrac{14}{5} \bar\psi^i (\gamh^j \Dgen_{ij} \rho
         - \tfrac15 \gamh_i \gamh^{jj'} \Dgen_{jj'} \rho) \\
      &-\tfrac{14}{5} (\bar\rho\gamh^i \Dgen_{ij} \psi^j)
         -\tfrac{42}{25} (\bar\rho \gamh^{ij} \Dgen_{ij} \rho)
         \Big) ,
\end{aligned}
\end{equation}
where we use the $\Spin(5)$ covariant spinor conjugate (see appendix~\ref{app:conv-d}). It is also important to note that there are two sets of suppressed indices on the spinors in this expression. These are the $\SU(2)$ indices for the five-dimensional symplectic Majorana spinors and the external $\Spin(6,1)$ indices, which must be summed over. For full details of the spinor conventions used, see appendices~\ref{app:cliff4} and~\ref{app:spin-decomp-4d}.

We have now rewritten all of the supergravity equations with manifest
$\Spin(5)$ symmetry following the prescription of
section~\ref{sec:abstract-sugra}.


\subsection{$d=7$ and $\tilde{H}_7=\SU(8)$}
\label{sec:7d}


\subsubsection{$\E7\times\bbR^+$ generalised geometry}

We follow the standard approach~\cite{deWN} of describing $E_{7(7)}$ in
terms of its $\SL(8,\bbR)$ subgroup, following the notation
of~\cite{PW}\footnote{Note however that when it comes to spinors, here
   we take instead $\gamma^{(7)} = -\ii$, the opposite choice to that
   in~\cite{PW}, and we also use a different normalisation of
   our $\SU(8)$ indices.}.     
We denote indices transforming under $\SL(8,\bbR)$ by
$i,j,k,\dots=1,\dots,8$. 

Generalised vectors transform in the $\rep{56}$ representation of
$\E7$, which under $\SL(8,\bbR)$ decomposes into the sum
$\rep{28}+\rep{28}'$ of bivectors and two-forms. We can introduce a
basis $\{\hat{E}_{ii'},\check{E}^{ii'}\}$ transforming under $\E7$ and
write a generalised vector as
\begin{equation}
   \label{eq:4}
   V = \tfrac{1}{2} V^{ii'} \hat{E}_{ii'} 
         + \tfrac{1}{2} \tilde{V}_{ii'} \check{E}^{ii'} .
\end{equation}
In the conformal split frame~\eqref{eq:geom-basis}, we can identify 
\begin{equation}
\begin{aligned}
   V^{a8} &= v^a , & && && 
   V^{ab} &= \tfrac{1}{5!} 
      \epsilon^{abc_1\dots c_5} \sigma_{c_1\dots c_5} , \\
   \tilde{V}_{a8} &= \tfrac{1}{7!} 
      \epsilon^{b_1\dots b_7} \tau_{a,b_1\dots b_7} , & && && 
   \tilde{V}_{ab} &= \omega_{ab} , 
\end{aligned}
\end{equation}
where $v^a, \omega_{ab},$ etc. are as in~\eqref{eq:split-iso}, with
the obvious corresponding identification of $\hat{E}_{a8}$ etc. The
partial derivative $\der_\mu$ is lifted into $E^*$, with a conformal
factor due to the form of the conformal split frame, as
\begin{equation}
\begin{gathered}
   \der_{a8\ph{'}} = \tfrac12 \ee^\Delta \der_a , \qquad \qquad
   \der_{ab\ph{'}} = 0 , \qquad \qquad
   \tilde\der^{ii'} = 0 .
\end{gathered}
\end{equation}
In this notation, the Dorfman
derivative~\eqref{eq:Lgen-cov}, the antisymmetrisation of which is the
``exceptional Courant bracket'' of~\cite{PW}, can then be written in
the coordinate frame~\eqref{eq:coord} as 
\begin{equation}
\begin{aligned}
   (\Lgen_V W)^{ii'} &= 
      V^{jj'}\der_{jj'}W^{ii'} + 4 W^{j[i} \der_{jj'} V^{i']j'} 
      \\ & \qquad 
      + W^{ii'}\der_{jj'}V^{jj'} 
      - \tfrac{1}{4}\epsilon^{ii'jj'kk'll'}\tilde{W}_{jj'}
         \der_{kk'}\tilde{V}_{ll'}  , \\
   (\Lgen_V W)_{ii'} &= 
       V^{jj'} \der_{jj'} \tilde{W}_{ii'} 
       - 4 \tilde{W}_{j[i}\der_{i']j'}V^{jj'} 
       - 6 W^{jj'} \der_{[jj'} \tilde{V}_{ii']} ,
\end{aligned}
\end{equation}
where $\epsilon^{i_1\dots i_8}$ is the totally antisymmetric symbol
preserved by $\SL(8,\bbR)$. 

A generic $\E7\times\bbR^+$ generalised connection
$\Dgen=(\Dgen_{ii'},\tilde{D}^{ii'})$ acting on $V\in\Gs{E}$ takes the
form
\begin{equation}
\begin{aligned}
   D_{ii'} V^{jj'} &= \der_{ii'} V^{jj'} 
      + \Omega_{ii'\ph{j}k}^{\ph{ii'}j} V^{kj'}
      + \Omega_{ii'\ph{j'}k}^{\ph{ii'}j'} V^{jk}
      + *\Omega_{ii'}^{\ph{ii'}jj'kk'} \tilde{V}_{kk'} , \\
   D_{ii'} \tilde{V}_{jj'} &= \der_{ii'} \tilde{V}_{jj'} 
      - \Omega_{ii'\ph{k}j}^{\ph{ii'}k} \tilde{V}_{kj'}
      - \Omega_{ii'\ph{k}j'}^{\ph{ii'}k} \tilde{V}_{jk}
      + \Omega_{ii'\,jj'kk'} V^{kk'} , \\
   \tilde{D}^{ii'} V^{jj'} &= \tilde{\der}^{ii'} V^{jj'} 
      + \tilde{\Omega}^{ii'\,j}_{\ph{ii'\,j}k} V^{kj'}
      + \tilde{\Omega}^{ii'\,j'}_{\ph{ii'\,j'}k} V^{jk}
      + *\tilde{\Omega}^{ii'\,jj'kk'} \tilde{V}_{kk'} , \\
   \tilde{D}^{ii'} \tilde{V}_{jj'} &= \tilde{\der}^{ii'} \tilde{V}_{jj'} 
      - \tilde{\Omega}^{ii'\,k}_{\ph{ii'\,k}j} \tilde{V}_{kj'}
      - \tilde{\Omega}^{ii'\,k}_{\ph{ii'\,k}j'} \tilde{V}_{jk}
      + \tilde{\Omega}^{ii'}_{\ph{ii'}jj'kk'} V^{kk'} ,
\end{aligned}
\end{equation}
where
$*\Omega_{ii'}^{\ph{ii'}jj'kk'}=\frac{1}{4}\epsilon^{jj'kk'll'mm'}
\Omega_{ii'\,ll'mm'}$ and similarly for $*\tilde{\Omega}^{ii'jj'kk'}$.


\subsubsection{$\SU(8)$ structures and supergravity}
\label{sec:7d-expr}

In seven dimensions $\Hd=\SU(8)/\bbZ_2$ and the common subgroup of
$\Hd$ and the $\SL(8,\bbR)$ subgroup that we used to define $\E7$ is
$SO(8)$. We define the sub-bundle $P\subset\tilde{F}$ of
$\SU(8)/\bbZ_2$ frames as the set of frames where the generalised
metric~\eqref{eq:flat-gen-metric} can be written as  
\begin{equation}
\label{eq:7d-gen-metric}
   G(V, W) = \tfrac12 \big( 
      \delta_{ij} \delta_{i'j'} V^{ii'} W^{jj'} 
      + \delta^{ij} \delta^{i'j'} \tV_{ii'} \tilde W_{jj'} \big) ,
\end{equation}
where $\delta_{ij}$ is the flat $\SO(8)$ metric. To write sections of
$E$ with manifest $\SU(8)$ indices $\alpha, \beta, \gamma, \dots
\ph{,} \!  \! \!  = 1, \dots, 8$ one uses the $SO(8)$ gamma matrices
\begin{equation}
\begin{aligned}
   V^{\alpha \beta} &=  \ii (\gamh_{ij})^{\alpha \beta} 
      \big(V^{ij} + \ii \tV^{ij }\big) ,\\
   \bar{V}_{\alpha \beta} &= - \ii (\gamh^{ij})_{\alpha \beta} 
      \big( V_{ij} - \ii \tV_{ij }\big) .
\end{aligned}
\end{equation}
where, $\gamh^{ij}$ are defined in~\eqref{eq:gamma8} and, when restricted to the $\Spin(8)$ subgroup $\alpha, \beta, \dots$ indices are raised and lowered using the intertwiner
$\tilde{C}$ (see appendix~\ref{app:Intertwiners}).

The eleven-dimensional supergravity fermion fields can be decomposed
into complex seven-dimensional spinors following the discussion
in~\ref{app:spin-decomp-7d}. Using the embedding 
$\Spin(7)\subset\Spin(8)\subset\SU(8)$, discussed in detail in
appendix~\ref{app:cliff7}, they can be identified as
$\SU(8)$ representations as follows. For the spinors we simply have
\begin{equation}
\begin{aligned}
   \varepsilon^{\alpha} 
      &= \ee^{-\Delta/2} (\varepsilon^{\text{sugra}})^{\alpha} 
      &\in \Gs{S^-},\\
   \bar{\rho}_\alpha 
      &= \ii \ee^{\Delta/2} \tilde{C}_{\alpha \beta}
          (\gamma^{(7)} \rho^{\text{sugra}})^{\beta}  
      &\in \Gs{S^+}.
\end{aligned}
\end{equation}
Note the need to include the conformal factors in the
definitions and also that, though we write $\bar\rho$ since it is embedded into the $\rep{\bar{8}}$ representation of $\SU(8)$, $\bar\rho_\alpha$ is defined in terms of the un-conjugated $ \rho^{\text{sugra}}$. The $\rep{8}$ and $\rep{\bar{8}}$ representations are simply the
fundamental and anti-fundamental so are related by
conjugation so that $\bar{\varepsilon}_\alpha=(\varepsilon^\beta)^*A_{\dot\beta\alpha}$, using
the $\SU(8)$-invariant intertwiner $A$ (see appendix~\ref{app:Intertwiners}).

For the 56-dimensional vector-spinor we proceed in two
steps, first embedding into $\Spin(8)$ by writing
\begin{equation}
\begin{aligned}	
   \psi_{a8}^{\scriptscriptstyle\Spin(8)} 
      &= \tfrac14 \ee^{\Delta/2}\big( 
          \delta^b{}_a + \tfrac12 \gamma_a \gamma^b \big)
          \psi^{\text{sugra}}_b ,\\
   \psi_{ab}^{\scriptscriptstyle\Spin(8)} 
      &=  - \tfrac12\ee^{\Delta/2} \gamma^{(7)} \big(  
          \gamma_{[a} \delta_{b]}{}^c 
          - \tfrac14 \gamma_{ab} \gamma^c  \big)\psi^{\text{sugra}}_c ,
\end{aligned}
\end{equation}
and then into $\SU(8)$ as
\begin{equation}
\begin{aligned}
   \psi^{\alpha \beta \gamma} 
       &= \tfrac{1}{3}\ii (\gamh^{ii'})^{[\alpha \beta} 
          (\psi_{ii'}^{\scriptscriptstyle\Spin(8)})^{\gamma]}  
          \in \Gs{J^-}.\\
\end{aligned}
\end{equation}

A generalised connection is compatible with the generalised
metric~\eqref{eq:7d-gen-metric} if $\Dgen G=0$. For such connections,
we can define the generalised spinor derivative via the adjoint action
of $\SU(8)$ given in~\cite{PW}. Acting on $\chi\in\Gs{S^-}$ we have 
\begin{equation}
\begin{aligned}
   \Dgen_{ii'} \chi &= \der_{ii'} \chi \,
      + \tfrac14 \Omega_{ii'jj'} \gamh^{jj'} \chi
      - \tfrac{1}{48}\ii \Omega_{ii' k_1 \dots k_4} 
          \gamh^{k_1 \dots k_4} \chi ,\\
   \tilde\Dgen_{ii'} \chi &= \tilde\der_{ii'} \chi
      +\tfrac14 \tilde{\Omega}_{ii'jj'} \gamh^{jj'} \chi
      - \tfrac{1}{48}\ii \tilde{\Omega}_{ii'k_1 \dots k_4} 
          \gamh^{k_1 \dots k_4} \chi .
\end{aligned}
\end{equation}
where we have used the $\SO(8)$ metric $\delta_{ij}$ to lower
indices. An example of such a generalised connection is the
one~\eqref{eq:Dnabla} defined by the Levi--Civita connection $\LC$
\begin{equation}
\begin{aligned}
   \Dgen^\LC_{ii'} \chi 
      &= \begin{cases}
         \tfrac{1}{2} \ee^\Delta \left(
            \der_{a} + \tfrac14 \omega_{a\,bc} \gamh^{bc} \right) \chi 
         & \text{if $i=a$ and $i'=8$} \\
         0 & \text{if $i=a$ and $i'=b$} 
      \end{cases} ,\\
   \tilde\Dgen^\LC_{ii'} \chi &= 0.
\end{aligned}
\end{equation}
where $\omega_{a\,bc}$ is the usual spin-connection. 

We can construct a torsion-free compatible connection $\Dgen$, by
shifting $\Dgen^\LC$ by an additional connection piece
$\Sigma$, such that its action on $\chi\in\Gs{S^-}$ is
given by 
\begin{equation}
\begin{aligned}
   \Dgen_{ii'} \chi &= \Dgen^\nabla_{ii'} \chi \,
      + \tfrac14 \Sigma_{ii'jj'} \gamh^{jj'} \chi
      - \tfrac{1}{48}\ii \Sigma_{ii' k_1 \dots k_4} 
         \gamh^{k_1 \dots k_4} \chi ,\\
   \tilde\Dgen_{ii'} \chi &= \tilde\Dgen^\nabla_{ii'} \chi
      + \tfrac14 \tilde{\Sigma}_{ii'jj'} \gamh^{jj'} \chi
      - \tfrac{1}{48}\ii \tilde{\Sigma}_{ii'k_1 \dots k_4} 
         \gamh^{k_1 \dots k_4} \chi .
\end{aligned}
\end{equation}
where, in the conformal split frame, 
\begin{equation}
\begin{aligned}
   \Sigma_{ii'jj'} &= -\tfrac13\ee^{\Delta} \delta_{ij}\tilde{K}_{i'j'} 
      + \tfrac{1}{42}\ee^{\Delta}\tF \delta_{ij}\delta_{i'j'}
      - \ee^{\Delta}\delta_{ij}\der_{i'j'}\Delta + \am_{ii'jj'}, \\
   \tilde{\Sigma}_{ii'jj'} &= \tfrac13 \ee^{\Delta}K_{ii'jj'}
      - \tfrac16 \ee^{\Delta}K_{jj'ii'} + \tilde{\am}_{ii'jj'}, \\
   \Sigma_{i_1\dots i_6} &= \am_{i_1\dots i_6}, \\
   \tilde{\Sigma}_{i_1\dots i_6} &= \tilde{\am}_{i_1\dots i_6} .
\end{aligned}
\end{equation}
In this expression primed and unprimed indices are antisymmetrised
implicitly, $(\am,\tilde{\am})$ are the undetermined
components\footnote{These are sections of the
   $\textbf{1280}+\bar{\textbf{1280}}$ representations of $\SU(8)$,
   see~\cite{CSW2}.}, $\tF = \tfrac{1}{7!} \epsilon^{a_1 \dots a_7}
\tF_{a_1 \dots a_7}$ and 
\begin{equation}
\begin{aligned}
   K_{ii'jj'} &= \begin{cases}
         (*F)_{abc} \, &\text{for } (i,i,'j,j')=(a,b,c,8) \\
         0 \, &\text{otherwise} 
      \end{cases}, \\
   \tilde{K}_{ij} &= \begin{cases} 
         \tF \, &\text{for } (i,j)=(8,8) \\
         0 \, &\text{otherwise} 
      \end{cases},
\end{aligned}
\end{equation}
give the embedding of the supergravity fluxes. The connection can be
rewritten in $\SU(8)$ indices through 
\begin{equation}
\begin{aligned}
   \Dgen^{\alpha \beta} &=  \ii (\gamh^{ij})^{\alpha \beta} 
      \big( \Dgen_{ij} + \ii \tilde\Dgen_{ij }\big) ,\\
   \bar\Dgen_{\alpha \beta} &= - \ii (\gamh_{ij})_{\alpha \beta} 
      \big( \Dgen^{ij} - \ii \tilde\Dgen^{ij }\big) .
\end{aligned}
\end{equation}

With these definitions, we can now give the explicit form of the
unique operators~\eqref{eq:unique-ops} in $\SU(8)$ indices 
\begin{equation}
\label{7d-proj}
\begin{aligned}
   (\DSJ \varepsilon)^{\alpha \beta \gamma} 
      &= \Dgen^{[\alpha \beta} \varepsilon^{\gamma]} , \\
   (\DSS  \varepsilon)_\alpha 
      &= -\bar\Dgen_{\alpha \beta} \varepsilon^\beta , \\
   (\DJJ \psi)_{\alpha \beta \gamma} 
      &= -\tfrac{1}{12} \, 
         \epsilon_{\alpha\beta\gamma\delta\delta'\theta_1\theta_2\theta_3} 
         \Dgen^{\delta \delta'} \psi^{\theta_1 \theta_2 \theta_3}, \\
   (\DJS \psi)^\alpha 
      &= \tfrac{1}{2} \bar\Dgen_{\beta \gamma} \psi^{\alpha \beta \gamma} ,
\end{aligned}
\end{equation}
where $\epsilon_{\alpha_1\dots\alpha_8}$ is the totally antisymmetric
symbol preserved by $\SU(8)$. 

From the first two we can immediately read off the supersymmetry
variations of the fermions~\eqref{eq:gen-susy-fermions} 
\begin{align}
   \delta \psi^{\alpha \beta \gamma} 
      &= \Dgen^{[\alpha \beta} \varepsilon^{\gamma]} , & 
   \delta \bar{\rho}_\alpha &= -\bar\Dgen_{\alpha \beta} \varepsilon^\beta ,
\end{align}
while the variations of the bosons~\eqref{eq:gen-susy-bosons} can be packaged as
\begin{equation}
   \delta G_{AB} 
      = \begin{pmatrix} 
         \delta G_{\alpha \beta \gamma \delta} & 
         \delta G_{\alpha \beta}{}^{\gamma \delta} \\
         \delta G^{\alpha \beta}{}_{\gamma \delta} 
         & \delta G^{\alpha \beta \gamma \delta} \end{pmatrix} 
      =  \frac{1}{\volG}\begin{pmatrix} 
         \delta\bar{ H}_{\alpha \beta \gamma \delta} & 0 \\
         0 & \delta H^{\alpha \beta \gamma \delta} \end{pmatrix} 
         - G_{AB}\; \delta \log \volG 
\end{equation}
with
\begin{equation}
\begin{aligned}
   \delta H^{\alpha \beta \gamma \delta} 
      &= -\tfrac{3}{16} \big( 
         \varepsilon^{[\alpha} \psi^{\beta \gamma \delta]}
         + \tfrac{1}{4!} 
            \epsilon^{\alpha\beta\gamma\delta\alpha'\beta'\gamma'\delta'}
            \bar\varepsilon_{\alpha'} \bar\psi_{\beta' \gamma' \delta'} 
            \big) \\
   \delta \log \volG 
      &= \bar{\rho}_\alpha \varepsilon^\alpha 
         + \rho^\alpha \bar\varepsilon_\alpha 
\end{aligned}
\end{equation}

The fermion equations of motion~\eqref{eq:gen-fermions} are
\begin{equation}
\begin{aligned}
   - \tfrac{1}{12} \, 
      \epsilon_{\alpha\beta\gamma\delta\delta'\theta_1\theta_2\theta_3} 
      \Dgen^{\delta \delta'} \psi^{\theta_1 \theta_2 \theta_3}
   + 2 \bar\Dgen_{[\alpha \beta} \bar{\rho}_{\gamma]} &= 0 ,\\
   \Dgen^{\alpha \beta} \bar{\rho}_\beta 
      - \tfrac{1}{2} \bar\Dgen_{\beta \gamma} \psi^{\alpha \beta \gamma} &= 0 .
\end{aligned}
\end{equation}
As before, the curvatures can be obtained by taking the supersymmetry variations of the fermion equations of motion and after some algebra one obtains the expressions
\begin{equation}
\begin{aligned}
   \GenRic_{\alpha \beta \gamma \delta} \varepsilon^\delta 
      &= -2 \big( 
         \bar\Dgen_{[\alpha \beta} \bar\Dgen_{\gamma \delta]}
         + \tfrac{1}{4!} \, 
            \epsilon_{\alpha\beta\gamma\delta\epsilon\epsilon'\theta\theta'} 
            \Dgen^{\epsilon \epsilon'} \Dgen^{\theta \theta'} 
            \big)\varepsilon^\delta
         - \BLie{\bar\Dgen_{[\alpha \beta}}{\bar\Dgen_{\gamma] \delta}} 
            \varepsilon^\delta , \\
   \tfrac16 \GenS\varepsilon^\alpha 
      &= - \tfrac23 \left( \{ \Dgen^{\alpha \gamma}, \bar\Dgen_{\beta \gamma} \}
            - \tfrac18 \delta^\alpha {}_{\beta} 
               \{ \Dgen^{\gamma \delta}, \bar\Dgen_{\gamma \delta} \}
            \right) \varepsilon^\beta 
      \\ & \qquad \qquad 
         - \tfrac13 \left( 
            [ \Dgen^{\alpha \gamma},  \bar\Dgen_{\beta \gamma} ]
               - \tfrac18 \delta^\alpha {}_{\beta} 
               [ \Dgen^{\gamma \delta},  \bar\Dgen_{\gamma \delta}]
            \right) \varepsilon^\beta
	 - \tfrac18 \BLie{\Dgen^{\beta\gamma}}{\bar\Dgen_{\beta \gamma}}
            \varepsilon^{\alpha} .
\end{aligned}
\end{equation}
The vanishing of these then corresponds to the bosonic equations of motion~\eqref{eq:gen-bosons}. As for $d=4$, we again observe that the curvatures contain terms symmetric in the two connections, in the representations identified in~\cite{CSW2}. 

The bosonic action~\eqref{eq:S-gen} takes the form
\begin{equation}
   S_{\text{B}} =  \frac{1}{2\kappa^2} \int \volG \GenS ,
\end{equation}
while the fermion action~\eqref{eq:gen-fermion-action} is
\begin{equation}
\label{eq:su8-fermion-action}
\begin{aligned}
   S_{\text{F}} = \frac{3}{2\kappa^2}   \int \volG \Big( 
      & \tfrac{1}{4!}
         \epsilon_{\alpha_1\alpha_2\alpha_3\beta\beta'\gamma_1\gamma_2\gamma_3}
           \psi^{\alpha_1\alpha_2\alpha_3} \Dgen^{\beta\beta'} 
           \psi^{\gamma_1\gamma_2\gamma_3}  
      \\ & \qquad 
        + \bar{\rho}_\alpha \bar\Dgen_{\beta \gamma} 
           \psi^{\alpha\beta\gamma} 
        - \psi^{\alpha \beta \gamma} 
           \bar\Dgen_{\alpha\beta} \bar{\rho}_{\gamma}
        - 2 \bar{\rho}_\alpha \Dgen^{\alpha \beta} \bar{\rho}_\beta
        + \text{cc} \Big).
\end{aligned}
\end{equation}
This completes the rewriting of the seven-dimensional theory with
explicit local $\SU(8)$ symmetry following from the natural
generalised geometry construction of
section~\ref{sec:abstract-sugra}. 


\section{Conclusions and discussion}
\label{sec:conc}


As promised at the end of~\cite{CSW2} we have provided a reformulation
of eleven-dimensional supergravity, including the fermions to leading
order, such that its larger bosonic symmetries are manifest. This was
accomplished by writing down an analogue of Einstein gravity for $\Edd
\times \bbR^+$ generalised geometry, the fermion fields embedding
directly into representations of the local symmetry group
$\dHd$. To summarise, the supergravity is described by a simple
set of equations which are manifestly diffeomorphism, gauge and
$\dHd$-covariant 
\begin{equation}
\begin{aligned}
   \text{Equations of Motion} \quad & && && && 
      \qquad \text{Supersymmetry} \\*[0.3em]
   \left\{ \begin{aligned}
      \DJJ \psi  
         + \tfrac{11-d}{9-d}\DSJ \rho &= 0, \\
      \DJS \psi  
         + \DSS \rho &= 0, \\
      \GenRicci_{AB} &= 0,
   \end{aligned} \right. \; & && && &&
   \left\{ \begin{aligned}
      \delta \psi &= \DSJ \varepsilon  , \\
      \delta \rho &= \DSS \varepsilon  , \\
      \delta G &= (\psi \proj{\adj{P}^\perp} 
              \varepsilon )
           + (\rho \proj{\adj{P}^\perp} \varepsilon ) .
   \end{aligned} \right. 
\end{aligned}
\end{equation}
This reinforces the conclusion of~\cite{CSW1} that generalised
geometry is a natural framework with which to formulate supergravity. 

It is important to note that these equations are equally applicable to
the reformulation of ten-dimensional type IIA or IIB supergravity
restricted to warped products of Minkowski space and a
$d-1$-dimensional manifold, where all the bosonic degrees (NSNS and
RR) are unified into the generalised metric $G$. Matching to the
familiar forms of the type II supergravity requires identifying the
appropriate $\GL(d-1,\bbR)$ and $\Spin(d-1)$ subgroups of $\Edd$ and
$\dHd$ respectively\footnote{Some of the details are given in
   appendix~B of~\cite{CSW2}.}. Alternatively by identifying the
appropriate $O(d-1,d-1)$ subgroup one can decompose to match to the
$O(d-1,d-1)\times\bbR^+$ generalised geometrical description
of~\cite{CSW1}. 

A surprising outcome of our work is the observation that, despite the
fact that the geometric construction is entirely bosonic,
supersymmetry is deeply integrated in the formalism -- torsion-free,
metric-compatible connections describe the variation of the fermions
and the equations of motion of the fermions close under supersymmetry
on the bosonic generalised curvatures. This relation between
generalised geometry and supersymmetry is clearly something that
warrants further exploration. One might try to formulate other
supergravities, such as six-dimensional $N=(1,0)$, which should
provide further evidence of this connection. It is of course also
interesting to see how one might extend the generalised geometry to
make supersymmetry manifest.

One problem that generalised geometry is particularly well suited to
tackle is that of describing generic supersymmetric vacua with 
flux~\cite{GLSW,GO1,GO2}. It turns out that the
Killing spinor equations can be shown to be equivalent to
integrability conditions on the generalised connection $\Dgen$. One
then expects that its special holonomy $G\subset\dHd$ can be
used to classify flux backgrounds. The language we developed in
section~\ref{sec:7d} will be especially useful for studying
compactifications of eleven-dimensional supergravity down to
four-dimensional spacetime, something we will elaborate on in
upcoming work. 


\acknowledgments

We would like to thank Mariana Gra\~{n}a for helpful discussions. This
work was supported by the STFC grant ST/J000353/1. The work of C.~S-C.~has been
supported by an STFC PhD studentship and by the German Science
Foundation (DFG) under the Collaborative Research Center (SFB) 676
``Particles, Strings and the Early Universe''. A.~C.~is supported
by the Portuguese Funda\c c\~ao para a Ci\^encia e a Tecnologia under
grant SFRH/BD/43249/2008. D.~W.~also thanks the Isaac Newton Institute for
Mathematical Sciences at Cambridge University and the Simons Center
for Geometry and Physics at Stony Brook
 University for hospitality and
support during the completion of this work. 


\appendix


\section{Conventions in $d$ dimensions}
\label{app:conv-d}


\subsection{Tensor notation}
\label{app:tensor}

We use the indices $m,n,p, \dots $ as the coordinate indices and
$a,b,c \dots$ for the tangent space indices. We take symmetrisation of
indices with weight one. Given a polyvector $w\in\Lambda^pTM$ and a
form $\lambda\in\Lambda^qT^*M$, we write in components 
\begin{equation}
\begin{aligned}
   w &= \frac{1}{p!} w^{m_1 \dots m_p} 
      \frac{\der}{\der x^{m_1}} \wedge \dots 
          \wedge \frac{\der}{\der x^{m_p}} , \\ 
   \lambda &= \frac{1}{q!} \lambda_{m_1 \dots m_q} 
       \dd x^{m_1} \wedge \dots \wedge \dd x^{m_q} , 
\end{aligned}
\end{equation}
so that wedge products and contractions are given by 
\begin{equation}
\begin{aligned}
   \left(w \wedge w'\right)^{m_1\dots m_{p+p'}} 
       &= \frac{(p+p')!}{p!p'!}
          w^{[m_1 \dots m_p}u^{m_{p+1} \dots m_{p+p'}]} , \\
   \left(\lambda \wedge \lambda'\right)_{m_1\dots m_{q+q'}} 
       &= \frac{(q+q')!}{q!q'!}
          \lambda_{[m_1 \dots m_q} \mu_{m_{p+1} \dots m_{q+q'}]} , \\
   \left(w\inn \lambda\right)_{a_1\dots a_{q-p}} 
      &:= \frac{1}{p!} w^{c_1\dots c_p}
         \lambda_{c_1\dots c_p a_1\dots a_{q-p}} &&&& 
         \text{if $p\leq q$} , \\
   \left(w\inn \lambda\right)^{a_1\dots a_{p-q}} 
      &:= \frac{1}{q!} w^{a_1\dots a_{p-q}c_1\dots c_q}
         \lambda_{c_1\dots c_q} &&&& 
         \text{if $p \geq q$} .  
\end{aligned}
\end{equation}
Given the tensors $t\in TM\otimes\Lambda^7TM$, $\tau\in
T^*M\otimes\Lambda^7T^*M$ and $a\in TM\otimes T^*M$ with components 
\begin{equation}
\begin{aligned}
   t &= \frac{1}{7!} w^{m,m_1 \dots m_7} 
      \frac{\der}{\der x^m} \otimes
          \frac{\der}{\der x^{m_1}}\wedge \dots 
          \wedge \frac{\der}{\der x^{m_7}} , \\ 
   \tau &= \frac{1}{7!} \tau_{m,m_1 \dots m_7} 
       \dd x^m \otimes \dd x^{m_1} \wedge \dots \wedge \dd x^{m_q} , \\
   a &= a^m{}_n \frac{\der}{\der x^m}\otimes \dd x^n , 
\end{aligned}
\end{equation}
we also use the ``$j$-notation'' from~\cite{PW,CSW2}, defining 
\begin{equation}
\begin{aligned}
   \left(w\inn \tau\right)_{a_1\dots a_{8-p}}
      &:= \frac{1}{(p-1)!} w^{c_1\dots c_p} 
         \tau_{c_1,c_2\dots c_p a_1\dots a_{8-p}} , \\
   \left(t\inn\lambda\right)^{a_1\dots a_{8-q}}
      &:= \frac{1}{(q-1)!} t^{c_1,c_2\dots c_q a_1\dots a_{8-q}}
         \lambda_{c_1\dots c_q} , \\
   \left(t\inn \tau\right) 
      & := \frac{1}{7!} t^{a,b_1\dots b_7}
         \tau_{a,b_1\dots b_7} , \\
   \left(jw\wedge w'\right)^{a,a_1\dots a_7}
      &:= \frac{7!}{(p-1)!(8-p)!}
         w^{a[a_1\dots a_{p-1}}w^{\prime a_p\dots a_7]} , \\
   \left(j\lambda\wedge\lambda'\right)_{a,a_1\dots a_7}
      &:= \frac{7!}{(q-1)!(8-q)!}
         \lambda_{a[a_1\dots a_{q-1}}\lambda'_{a_q\dots a_7]} , \\
   \left(jw\inn j\lambda\right)^a{}_b
      &:= \frac{1}{(p-1)!} w^{ac_1\dots c_{p-1}}
         \lambda_{bc_1\dots c_{p-1}} ,\\
   \left(jt\inn j\tau\right)^a{}_b
      &:= \frac{1}{7!} t^{a,c_1\dots c_7}
         \tau_{b,c_1\dots c_7} . 
\end{aligned}
\end{equation}
%


\subsection{Metrics, connections and curvatures}
\label{app:conv}

The $d$ dimensional metric $g$ is always positive definite. We define
the orientation, $\epsilon_{1 \dots d} = \epsilon^{1 \dots d} = +1$,
and use the conventions
\begin{equation}
\label{eq:*norm}
\begin{aligned}
   * \lambda_{m_1\dots m_{d-q}} 
      &= \tfrac{1}{q!} \sqrt{|g|} 
         \epsilon_{m_1 \dots m_{d-k} n_1 \dots n_q} 
         \lambda^{n_1 \dots n_q} , \\
   \lambda^2 &= \lambda_{m_1\dots m_q}\lambda^{m_1\dots m_q} . 
\end{aligned}
\end{equation}
Let $\nabla_m v^n=\der_m v^n+\omega_m{}^n{}_p v^p$ be a general
connection on $TM$. The torsion $T\in\Gs{TM\otimes\Lambda^2T^*M}$ of
$\nabla$ is defined by 
\begin{equation}
\label{eq:Tdef}
   T(v,w) = \nabla_v w - \nabla_w v - \BLie{v}{w} . 
\end{equation}
or concretely, in coordinate indices, 
\begin{equation}
   T^m{}_{np} = \omega_n{}^m{}_p - \omega_p{}^m{}_n , 
\end{equation}
while, in a general basis where $v=v^a \hat{e}_a$ and $\nabla_m
v^a=\der_m v^a+\omega_m{}^a{}_bv^b$, one has  
\begin{equation}
\label{eq:Tcomp}
   T^a{}_{bc} = \omega_b{}^a{}_c - \omega_c{}^a{}_b
       + \BLie{\hat{e}_b}{\hat{e}_c}^a . 
\end{equation}
The curvature of a connection $\nabla$ is given by the Riemann tensor
$\Riem \in \Gs{\Lambda^2T^*M \otimes TM \otimes T^*M}$, defined by
$\Riem(u,v)w=[\LC_u,\LC_v]w-\LC_{[u,v]}w$, or in components  
\begin{equation}
\label{eq:Rdef}
 \Riem_{mn\phantom{p}q}^{\phantom{mn}p}w^{q} 
      = [ \nabla_m , \nabla_n ] w^p - T^q{}_{mn}\nabla_{q}w^p .
\end{equation}
The Ricci tensor is the trace of the Riemann curvature
\begin{align}
\label{eq:Ricdef}
   \Ric_{mn}=\Riem_{pm\phantom{p}n}^{\phantom{mn}p} .
\end{align}
Given a metric $g$ the Ricci scalar for a metric-compatible connection
is defined by
\begin{align}
\label{eq:Sdef}
   \Scalar=g^{mn}\Ric_{mn} .
\end{align}
The Levi-Civita connection is the unique connection that is both
torsion free ($T=0$) and metric-compatible ($\nabla g=0$). 


\section{Clifford algebras and spinors}


\subsection{Clifford algebras, involutions and $\dHd$}
\label{app:invol}

The real Clifford algebras $\Cliff(p,q;\bbR)$ are generated by gamma
matrices satisfying 
\begin{align}
   \left\{ \gamma^m , \gamma^n \right\} &= 2 g^{mn}, 
   &  \gamma^{m_1 \dots m_k} &= \gamma^{[m_1} \dots \gamma^{m_k]} ,
\end{align}
where $g$ is a $d$-dimensional metric of signature $(p,q)$. Here we
will be primarily interested in $\Cliff(d;\bbR)=\Cliff(d,0;\bbR)$ and
$\Cliff(d-1,1;\bbR)$. The top gamma matrix is defined as
\begin{equation}
   \gamma^{(d)} 
      = \tfrac{1}{d!} \epsilon_{m_1 \dots m_d} 
         \gamma^{m_1 \dots m_{d}}
      = \begin{cases} 
           \gamma^0 \gamma^1 \dots \gamma^{d-1} 
             & \text{for $\Cliff(d-1,1;\bbR)$} \\
           \gamma^1 \dots \gamma^d 
             & \text{for $\Cliff(d;\bbR)$} 
          \end{cases} ,
\end{equation}
and one has $\BLie{\gamma^{(d)}}{\gamma^m}=0$ if $d$ is odd, while
$\{\gamma^{(d)},\gamma^m\}=0$ if $d$ is even, and  
\begin{equation}
\label{eq:gd2}
   (\gamma^{(d)})^2 
      = \begin{cases} 
         1 & \text{if $p-q=0,1 \pmod 4$} \\
         -1 & \text{if $p-q=2,3 \pmod 4$}
      \end{cases}
\end{equation}
We also use Dirac slash notation with weight one so that for $\omega
\in \Gs{\Lambda^k T^*M}$ %
\begin{equation}
   \slashed{\omega} 
      = \tfrac{1}{k!} \omega_{m_1 \dots m_k} \gamma^{m_1 \dots m_k} .
\end{equation}

The real Clifford algebras are isomorphic to matrix algebras over
$\bbR$, $\bbC$ or the quaternions $\bbH$. These are listed in
table~\ref{tab:clifford}. Note that in odd dimensions the pair
$\{1,\gamma^{(d)}\}$ generate the centre of the algebra, which is
isomorphic to $\bbR\oplus\bbR$ if $p-q=1\pmod 4$ and $\bbC$ if
$p-q=3\pmod 4$. In the first case $\Cliff(p,q;\bbR)$ splits into two
pieces with $\gamma^{(d)}$ eigenvalues of $\pm 1$. In the second case
$\gamma^{(d)}$ plays the role of $\ii$ under the isomorphism with
$\GL(2^{[d/2]},\bbC)$. 
\begin{table}[htb]
\begin{center}
\begin{tabular}{ll}
   $p-q \pmod 8$ &  $\Cliff(p,q;\bbR)$ \\
   \hline
   0, 2 & $\GL(2^{d/2},\bbR)$ \\
   1 & $\GL(2^{[d/2]},\bbR)\oplus \GL(2^{[d/2]}, \bbR)$ \\
   3, 7 & $\GL(2^{[d/2]},\bbC)$ \\
   4, 6 & $\GL(2^{d/2-1},\bbH)$ \\
   5 & $\GL(2^{[d/2]-1},\bbH)\oplus\GL(2^{[d/2]-1},\bbH)$ \\
\end{tabular}
\end{center}
\caption{Real Clifford algebras}
\label{tab:clifford}
\end{table}

There are three involutions of the algebra given by 
\begin{equation}
\begin{aligned}
   \gamma^{m_1\dots m_k} &\mapsto (-)^k \gamma^{m_1\dots m_k} , \\
   \gamma^{m_1\dots m_k} &\mapsto \gamma^{m_k\dots m_1} , \\
   \gamma^{m_1\dots m_k} &\mapsto (-)^k \gamma^{m_k\dots m_1} , 
\end{aligned}
\end{equation}
usually called ``reflection'', ``reversal'' and ``Clifford
conjugation''. The first is an automorphism of the algebra, the other
two are anti-automorphisms. The reflection involution gives a grading
of $\Cliff(p,q;\bbR)=\Cliff^+(p,q;\bbR)\oplus\Cliff^-(p,q;\bbR)$ into
odd and even powers of $\gamma^m$. The group $\Spin(p,q)$ lies in
$\Cliff^+(p,q;\bbR)$.  

The involutions can be used to define other subgroups of the Clifford
algebra. In particular one has 
\begin{equation}
   \tilde{H}_{p,q} = \{ g\in\Cliff(p,q;\bbR) : g^tg=1 \}
\end{equation}
$g^t$ is the image of $g$ under the reversal involution. For the
corresponding Lie alegbra we require $a^t+a=0$, and so the algebra is
generated by elements in the negative eigenspace of the
involution. For $d\leq 8$, this is the set $\{\gamma^{mn},
\gamma^{mnp},\gamma^{m_1\dots m_6},\gamma^{m_1\dots m_7}\}$. We see
that the maximally compact subgroups $\dHd\subset\Edd$ are given by 
\begin{equation}
   \dHd = \tilde{H}_{d,0}
\end{equation}
for the $\Cliff(d;\bbR)$ algebras\footnote{Note that $\tilde{H}_{7,0}$
   is strictly $U(8)$. Dropping the $\gamma^{(7)}$ generator one gets
   $\tilde{H}_7=\SU(8)$.}.  


\subsection{Representations of $\Cliff(p,q;\bbR)$ and intertwiners}
\label{app:Intertwiners}

It is usual to consider irreducible complex representations of
the gamma matrices acting on spinors. When $d$ is even there is only
one such representation. There are then three intertwiners realising
the involutions discussed above, namely, 
\begin{equation}
\begin{aligned}
   \gamma_{(d)} \gamma^m \gamma_{(d)}^{-1} &= - \gamma^m , \\
   C \gamma^m C^{-1} &= (\gamma^m)^T , \\
   \tilde{C} \gamma^m \tilde{C}^{-1} &= - (\gamma^m)^T , \\
\end{aligned}
\end{equation}
where $\tilde{C}=C\gamma^{(d)}$. There are four further intertwiners,
not all independent, giving
\begin{equation}
\begin{aligned}
   A \gamma^m A^{-1} &= (\gamma^m)^\dag , & && &&
   D \gamma^m D^{-1} &= (\gamma^m)^* , \\
   \tilde{A} \gamma^m \tilde{A}^{-1} &= - (\gamma^m)^\dag , & && && 
   \tilde{D} \gamma^m \tilde{D}^{-1} &= - (\gamma^m)^* .
\end{aligned}
\end{equation}
By construction we see that $\dHd$ is the group preserving $C$. 

When $d$ is odd there are two inequivalent irreducible
representations with either $\gamma^{(d)}=\pm1$ when $p-q=1\pmod 4$ or
$\gamma^{(d)}=\pm\ii$ when $p-q=3\pmod4$. Since here $\gamma^{(d)}$ is
odd under the reflection, this involution exchanges the two
representations. Thus only half of the possible intertwiners exist on
each. One has 
\begin{equation}
\begin{aligned}
   C \gamma^m C^{-1} &= (\gamma^m)^T , 
      && && && \text{if $d=1\pmod 4$} , \\
  \tilde{C} \gamma^m \tilde{C}^{-1} &= - (\gamma^m)^T , 
      && && && \text{if $d=3\pmod 4$} .
\end{aligned}
\end{equation}
while 
\begin{equation}
\begin{aligned}
   A \gamma^m A^{-1} &= (\gamma^m)^\dag , 
      && && && \text{if $p$ is odd} , \\
  \tilde{A} \gamma^m \tilde{A}^{-1} &= - (\gamma^m)^\dag , 
      && && && \text{if $p$ is even} , \\
   D \gamma^m D^{-1} &= (\gamma^m)^* , 
      && && && \text{if $p-q=1\pmod 4$} , \\
  \tilde{D} \gamma^m \tilde{D}^{-1} &= - (\gamma^m)^* , 
      && && && \text{if $p-q=3\pmod 4$} .
\end{aligned}
\end{equation}

Note that under reversal
$(\gamma^{(d)})^t=(-)^{d(d-1)/2}\gamma^{(d)}$ so when $d=3\pmod 4$ the
involution exchanges representations and we have no $C$
intertwiner. In particular for $\Cliff(d;\bbR)$ it maps
$\gamma^{(d)}=\ii$ to $\gamma^{(d)}=-\ii$. However, this map can also
be realised on each representation separately by the adjoint
$A\gamma^mA^{-1}=(\gamma^m)^\dag$. Hence for $d=3\pmod 4$ we can
instead define $\dHd$ as the group preserving $A$.

The conjugate intertwiners allow us to define Majorana and symplectic
Majorana representations when there is an isomorphism to real and
quaternionic matrix algebras respectively. Thus when $p-q=0,1,2\pmod
8$ one has $DD^*=1$ and can define a reality condition on the spinors
\begin{equation}
   \chi = (D\chi)^* .
\end{equation}
When $p-q=4,5,6\pmod 8$ one has $DD^*=-1$ one can define a symplectic
reality condition. Introducing a pair of $\SU(2)$ indices $A,B,\dots
\ph{,} \! \! = 1,2$ on the spinors with the convention for raising and
lowering these indices
\begin{align}
   \chi_A &= \epsilon_{AB} \chi^B , &
   \chi^A &= \epsilon^{AB} \chi_B , 
\end{align}
the symplectic Majorana condition is
\begin{equation}
   \eta^A = \epsilon^{AB} (D \eta^B)^* . 
\end{equation}
Note that for $p-q=0,6,7\pmod 8$ and $p-q=2,3,4\pmod 8$ one can
also define analogous Majorana and symplectic Majorana conditions
respectively using $\tilde{D}$. 


\subsubsection{$\Cliff(10,1;\bbR)$}

For $\Cliff(10,1;\bbR)\simeq\GL(32,\bbR)\oplus\GL(32,\bbR)$, following
the conventions of~\cite{GP} we take the representation with  
\begin{equation}
   \Gamma^{(11)} = \Gamma^0 \Gamma^1 \dots \Gamma^{10}= -1.
\end{equation}
The $D$ intertwiner defines Majorana spinors, while
$\tilde{C}=-\tilde{C}^T$ defines the conjugate
\begin{equation}
\label{eq:11d-conj}
   \varepsilon = (D \varepsilon)^* , \qquad \qquad 
   \bar\varepsilon = \varepsilon^T \tilde{C} .
\end{equation}
such that 
\begin{equation}
   \overline{\Gamma^{M_1 \dots M_k} \varepsilon} 
	= (-1)^{[(k+1)/2]} 
           \bar\varepsilon \Gamma^{M_1 \dots M_k} .
\end{equation}
%


\subsubsection{$\Cliff(4 ; \bbR)$ and $\Spin(5)$}
\label{app:cliff4}

For $\Cliff(4;\bbR)\simeq\GL(2,\bbH)$, $D^*D = - 1$ and we can use this to
introduce symplectic Majorana spinors, while we use $\tilde{C}$ to define the
conjugate spinor
\begin{equation}
   \chi^A = \epsilon^{AB} (D \chi^B)^* , \qquad
   \bar{\chi}_A = \epsilon_{AB} (\chi^B)^T \tilde{C}
\end{equation}
The other intertwiner $C=\tilde{C} \gamma^{(4)}$ provides a symplectic inner product on spinors, which
is preserved by $\{ \gamma^{mn}, \gamma^{mnp} \}$, i.e. the
$\tilde{H}_4 \cong \Spin(5)$ algebra. The $\Spin(5)$ gamma matrix
algebra can be realised explicitly by setting
\begin{equation}
   \gamh^{i} = \begin{cases} 
                 \gamma^{a} & i=a \\
		 \gamma^{(4)} & i = 5 
              \end{cases} ,
\end{equation}
and identifying $\gamma^{mnp}=-\epsilon^{mnpq} \gamma_q \gamma^{(4)}$. The same gamma
matrices give a representation of $\Cliff(5; \bbR)$ (with
$\gamma^{(5)} = +1$).   


\subsubsection{$\Cliff(7 ; \bbR)$ and $\Spin(8)$}
\label{app:cliff7}

For $\Cliff(7;\bbR)$ we take the representation with
$\gamma^{(7)}=-\ii$ and define conjugate spinors  
\begin{equation}
  \bar\varepsilon = \varepsilon^\dag A .
\end{equation}
This provides a hermitian inner product on spinors, which is preserved
by $\tilde{H}_7 \cong \SU(8)$, generated by $\{ \gamma^{mn},
\gamma^{mnp}, \gamma^{m_1 \dots m_6} \}$. The intertwiner
$\tilde{C}=\tilde{C}^T$ is preserved by a $\Spin(8)\subset\SU(8)$
subgroup. The corresponding generators can be written as 
\begin{equation}
\label{eq:gamma8}
   \gamh^{ij} = \begin{cases} 
          \gamma^{ab} & i=a , j=b \\
          +\gamma^{a} \gamma^{(7)} & i=a, j=8 \\
          -\gamma^b \gamma^{(7)} & i = 8, j=b 
       \end{cases} ,
\end{equation}
This representation has negative chirality in the sense that 
\begin{equation}
   \gamh^{i_1 \dots i_8} = - \epsilon^{i_1 \dots i_8} .
\end{equation}
We have the useful completeness relations, reflecting $\SO(8)$ triality,
\begin{align*}
   \gamh^{ij}{}_{\alpha \beta} \gamh_{ij}{}^{\gamma \delta} 
      &= 16 \delta^{\gamma \delta}_{\alpha \beta} , &
   \gamh^{ij}{}_{\alpha \beta} \gamh_{kl}{}^{\alpha \beta} 
      &= 16 \delta^{ij}_{kl} ,
\end{align*}
where we have used $\tilde{C}$ to raise and lower spinor indices, and
Fierz identity, which also serves as our definition of
$\epsilon_{\alpha_1 \dots \alpha_8}$,  
\begin{equation}
   \tfrac{1}{4!} 
     \epsilon_{\alpha \alpha' \beta \beta' \gamma \gamma' \delta \delta'}
     \gamh^{ij \gamma \gamma'} \gamh^{kl \delta \delta'}
     = 2 \gamh^{[ij}{}_{[\alpha \alpha'} \gamh^{kl]}{}_{\beta \beta']}
        - \gamh^{ij}{}_{[\alpha \alpha'} \gamh^{kl}{}_{\beta \beta']} .
\end{equation}

Note that as a representation of the $\Spin(8)$ algebra we can impose
a reality condition on the spinors $\chi=(D\chi)^*$ using the
intertwiner $\tilde{D}$ with $\tilde{D}^* \tilde{D} = +1$. For such a
real spinor the two possible definitions of spinor conjugate coincide
$\bar\chi = \chi^T \tilde{C} = \chi^\dagger A$. In fact there exists a
$\GL(8,\bbR)$ family of purely imaginary bases of gamma matrices such
that $\tilde{D} = 1$ and $A = \tilde{C}$. In such a basis we have
$\bar\varepsilon=\varepsilon^\dag\tilde{C} = \varepsilon^\dag A$ for a
general spinor $\varepsilon=\chi_1+\ii\chi_2$. Many of our $\SU(8)$
equations are written under a $\Spin(8) = \SU(8) \cap \SL(8,\bbR)$
decomposition in such an imaginary basis, and thus it is natural to
raise and lower spinor indices with the $\Spin(8)$ invariant
$\tilde{C}$.


\section{Spinor Decompositions}
\label{app:spin-decomp}

\subsection{$(10,1) \rightarrow (6,1) + (4,0)$}
\label{app:spin-decomp-4d}

We can decompose the $\Cliff(10,1; \bbR)$ gamma matrices as
\begin{equation}
\label{eq:4d-gamma-split}
	\Gamma^\mu = \gamma^\mu \otimes \gamma^{(4)} ,
	\hspace{40pt}
	\Gamma^m = 1 \otimes \gamma^m ,
\end{equation}
and the eleven-dimensional intertwiners as
\begin{equation}
	\tilde{C} = \tilde{C}_{(6,1)} \otimes \tilde{C}_{(4)} ,
	\hspace{40pt}
	D = D_{(6,1)} \otimes D_{(4)} .
\end{equation}

Introducing a basis of seven dimensional symplectic Majorana spinors $\{ \eta^A_I \}$, we can then decompose a general eleven-dimensional Majorana spinor as
\begin{equation}
	\varepsilon = \epsilon_{AB} \left( \eta^A_I \otimes \chi^{BI} \right) ,
\end{equation}
where $\{ \chi^{AI} \}$ are some collection of four dimensional symplectic Majorana spinors. All of the data of the eleven dimensional spinor is now contained in $\chi^{AI}$, the extra index $I$ serving as the external $\Spin(6,1)$ index.

The eleven dimensional spinor conjugate can be realised in terms of the four dimensional spinors $\chi^{AI}$ by setting
\begin{equation}
\label{eq:4d-spinor-conj}
	\bar\chi_{AI} = \epsilon_{AIBJ} (\chi^{BJ})^T \tilde{C}_{(4)} ,
\end{equation}
where $\epsilon_{AIBJ} = (\eta_{AI})^T \tilde{C}_{(6,1)} \eta_{BJ}$.

Clearly from the decomposition~\eqref{eq:4d-gamma-split} the action of the internal eleven dimensional gamma matrices is simply
\begin{equation}
	\Gamma^m \varepsilon \leftrightarrow \gamma^m \chi^{AI} ,
\end{equation}
and any eleven dimensional equation with only internal gamma matrices takes the same form in terms of $\chi^{AI}$. Thus, supressing the extra indices on $\chi$, the supergravity equations with fermions in section~\ref{sec:eom} take exactly the same form when written in terms of the four-dimensional spinors.

\subsection{$(10,1) \rightarrow (5,1) + (5,0)$}

We can use a complex decomposition of the $\Cliff(10,1; \bbR)$ gamma matrices as
\begin{equation}
	\Gamma^\mu = \gamma^\mu \otimes 1 ,
	\hspace{40pt}
	\Gamma^m = \gamma^{(6)} \otimes \gamma^m ,
\end{equation}
and the eleven-dimensional intertwiners as
\begin{equation}
	\tilde{C} = \tilde{C}_{(5,1)} \otimes C_{(5)} ,
	\hspace{40pt}
	D = D_{(5,1)} \otimes D_{(5)} .
\end{equation}
We introduce bases of positive and negative chirality symplectic Majorana-Weyl spinors $\{ \eta^{+A}_I\}$ and $\{ \eta^{-A}_I\}$ and decompose a general eleven-dimensional Majorana spinor as $\varepsilon = \varepsilon^+ + \varepsilon^-$ with
\begin{equation}
	\varepsilon^+ = \epsilon_{AB} \left( \eta^{+A}_I \otimes \chi^{IB}_1 \right)
	\hspace{40pt}
	\varepsilon^- =  \epsilon_{AB} \left( \eta^{-A}_I \otimes \chi^{IB}_2 \right) ,
\end{equation}
where $\{ \chi^{AI}_1 \}$ and $\{ \chi^{AI}_2 \}$ are two collections of five dimensional symplectic Majorana spinors, the extra $I$ indices serving as external $\Spin(5,1)$ indices.

The action of the internal eleven dimensional gamma matrices is then
\begin{equation}
	\Gamma^m \varepsilon^+ 
		= \epsilon_{AB} \left( \eta^{+A}_I \otimes \gamma^m \chi^{IB}_1 \right)
	\hspace{40pt}
	\Gamma^m \varepsilon^- 
		=  \epsilon_{AB} \left( \eta^{-A}_I \otimes (-\gamma^m) \chi^{IB}_2 \right) ,
\end{equation}
so that this indices a different representation of $\Cliff(5; \bbR)$ on each of $\chi_1$ and $\chi_2$. To see how to express eleven-dimensional spinor bilinears in this case, we expand
\begin{equation}
\begin{aligned}
	\bar\varepsilon \, \Gamma^{m_1 \dots m_k} \varepsilon' &=  
		\bar\varepsilon^+ \, \Gamma^{m_1 \dots m_k} \varepsilon'^-
		+ \bar\varepsilon^- \, \Gamma^{m_1 \dots m_k} \varepsilon'^+ \\
		&= \Big( (\eta^+_{AI})^T \tilde{C}_{(5,1)} \eta^-_{BJ} \Big)
			\Big( (\chi^{AI}_1)^T C_{(5)} (-1)^k \gamma^{m_1 \dots m_k} \chi'^{BJ}_2 \Big) \\
		& \qquad + \Big( (\eta^-_{AI})^T \tilde{C}_{(5,1)} \eta^+_{BJ} \Big)
			\Big( (\chi^{AI}_2)^T C_{(5)} \gamma^{m_1 \dots m_k} \chi'^{BJ}_1 \Big) ,
\end{aligned}
\end{equation}
so that we see that this pairs $\chi_1$ with $\chi'_2$ and $\chi_2$ with $\chi'_1$. We therefore define slightly different conjugates for $\chi_1$ and $\chi_2$ as
\begin{equation}
	\bar\chi_{1,AI} = \epsilon_{AIBJ} (\chi^{BJ}_1)^T C_{(5)},
	\hspace{40pt}
	\bar\chi_{2,AI} = \epsilon_{BJAI} (\chi^{BJ}_2)^T C_{(5)},
\end{equation}
where $\epsilon_{AIBJ} = (\eta^-_{AI})^T \tilde{C}_{(5,1)} \eta^+_{BJ}$. We can then suppress the extra indices and write
\begin{equation}
\label{eq:5d-bilinear-split}
	\bar\varepsilon \, \Gamma^{m_1 \dots m_k} \varepsilon'
		= \bar\chi_2 \gamma^{m_1 \dots m_k} \chi_1
			+ (-1)^k \bar\chi_1 \gamma^{m_1 \dots m_k} \chi_2 .
\end{equation}

Under this decomposition, an equation linear in fermions from section~\ref{sec:eom} becomes two copies of the same equation, one identical copy with ``$\chi_1$" and one copy with ``$\chi_2$" and the sign of the gamma matrices reversed. The fermion bilinears split into two terms as~\eqref{eq:5d-bilinear-split}.

\subsection{$(10,1) \rightarrow (4,1) + (6,0)$}

We make a complex decomposition of the $\Cliff(10,1; \bbR)$ gamma matrices as
\begin{equation}
\label{eq:6d-gamma-split}
	\Gamma^\mu = \ii\gamma^\mu \otimes \gamma^{(6)} ,
	\hspace{40pt}
	\Gamma^m = 1 \otimes \gamma^m ,
\end{equation}
and the eleven-dimensional intertwiners as
\begin{equation}
	\tilde{C} = C_{(4,1)} \otimes \tilde{C}_{(6)} ,
	\hspace{40pt}
	D = \tilde{D}_{(4,1)} \otimes D_{(6)}.
\end{equation}

Introducing a basis of five dimensional symplectic Majorana spinors $\{ \eta^A_I \}$ we can then decompose a general eleven-dimensional Majorana spinor as
\begin{equation}
	\varepsilon = \epsilon_{AB} \left( \eta^A_I \otimes \chi^{BI} \right) ,
\end{equation}
where $\{ \chi^{AI} \}$ are some collection of six dimensional symplectic Majorana spinors. All of the data of the eleven dimensional spinor is now contained in $\chi^{AI}$, the extra index $I$ serving as the external $\Spin(4,1)$ index.

The eleven dimensional spinor conjugate can be realised in terms of the six dimensional spinors $\chi^{AI}$ by setting
\begin{equation}
	\bar\chi_{AI} = \epsilon_{AIBJ} (\chi^{BJ})^T \tilde{C}_{(6)} ,
\end{equation}
where $\epsilon_{AIBJ} = -(\eta_{AI})^T C_{(4,1)} \eta_{BJ}$.

Clearly from the decomposition~\eqref{eq:4d-gamma-split} the action of the internal eleven dimensional gamma matrices is simply
\begin{equation}
	\Gamma^m \varepsilon \leftrightarrow \gamma^m \chi^{AI} ,
\end{equation}
and any eleven dimensional equation with only internal gamma matrices takes the same form in terms of $\chi^{AI}$. Thus, supressing the extra indices on $\chi$, the supergravity equations with fermions in section~\ref{sec:eom} take exactly the same form when written in terms of the six-dimensional spinors.

\subsection{$(10,1) \rightarrow (3,1) + (7,0)$}
\label{app:spin-decomp-7d}

We can use a complex decomposition of the $\Cliff(10,1; \bbR)$ gamma matrices as
\begin{equation}
\label{eq:7d-gamma-split}
	\Gamma^\mu = \gamma^\mu \otimes 1 ,
	\hspace{40pt}
	\Gamma^m = \ii\gamma^{(4)} \otimes \gamma^m ,
\end{equation}
and the eleven-dimensional intertwiners as
\begin{equation}
	\tilde{C} = \tilde{C}_{(3,1)} \otimes \tilde{C}_{(7)} ,
	\hspace{40pt}
	D = D_{(3,1)} \otimes \tilde{D}_{(7)} .
\end{equation}

We take a chiral decomposition of an eleven-dimensional Majorana spinor
\begin{equation}
	\varepsilon = \left( \eta^+_I \otimes \chi^I \right) 
	+ (D_{(3,1)} \eta^+_I)^*\otimes (\tilde{D}_{(7)} \chi^I)^* ,
\end{equation}
where $\gamma^{(4)} \eta^+_I = - \ii \eta^+_I$ so that $\{ \eta^+_I \}$ are a basis of complex Weyl spinors in the external space. The Majorana condition on $\varepsilon$ is automatic with no additional constraint on $\chi^I$, which is complex. Again the extra index $I$  on $\chi$ provides an external $\Spin(3,1)$ index.

The Clifford action of the internal eleven-dimensional gamma matrices then reduces to the action of the seven-dimensional gamma matrices on $\chi$
\begin{equation}
	\Gamma^m \varepsilon = 
		\eta^+_I \otimes (\gamma^m \chi^I)
		+ (D_{(3,1)} \eta^+_I)^*\otimes (\tilde{D}_{(7)} \gamma^m \chi^I)^* .
\end{equation}
To see how to write eleven-dimensional spinor bilinears in this language, we expand
\begin{equation}
\label{eq:7d-bilinear-split}
\begin{aligned}
	\bar\varepsilon \, \Gamma^{m_1 \dots m_k} \varepsilon' &= 
		\Big( (\eta^+_I)^T \tilde{C}_{(3,1)} \eta^+_J \Big)
			\Big( (\chi^I)^T \tilde{C}_{(7)} \gamma^{m_1 \dots m_k} \chi'^J \Big) \\
		& \qquad + \Big( (\eta^+_I)^T D^T_{(3,1)} \tilde{C}_{(3,1)} D_{(3,1)} \eta^+_J \Big)^*
			\Big( (\chi^I)^T \tilde{D}^T_{(7)} 
				\tilde{C}_{(7)} \tilde{D}_{(7)} \gamma^{m_1 \dots m_k} \chi'^J \Big)^* \\
		& = \Big( \bar\chi_I \gamma^{m_1 \dots m_k} \chi'^I \Big) 
			+ \Big( \text{cc} \Big) ,
\end{aligned}
\end{equation}
where we have made the definition
\begin{equation}
	\bar\chi_I = \epsilon_{IJ} (\chi^J)^T \tilde{C}_{(7)} ,
\end{equation}
with $\epsilon_{IJ} =  -(\eta^+_I)^T \tilde{C}_{(3,1)} \eta^+_J$. 

With these definitions, the equations linear in spinors in section~\ref{sec:eom} take the same form when written in terms of $\chi^I$, while the spinor bilinear expressions take the same form with a complex conjugate piece added to them.


\section{$\Edd\times\bbR^+$ and $\GL(d,\bbR)$}
\label{app:Edd}


In this appendix we review from~\cite{CSW2} the construction of
$\Edd\times\bbR^+$ in terms of $\GL(d,\bbR)$, the basic
representations and tensor products.

We will describe the action directly in terms of the bundles that
appear in the generalised geometry. We have 
\begin{equation}
\label{eq:Ed-bundles}
\begin{aligned}
   E &\simeq TM \oplus \Lambda^2T^*M \oplus \Lambda^5T^*M
         \oplus (T^*M \otimes\Lambda^7T^*M) , \\
   E^* &\simeq T^*M \oplus \Lambda^2TM \oplus \Lambda^5TM
         \oplus (TM \otimes\Lambda^7TM) , \\
   \adj{\tilde{F}} &\simeq \bbR \oplus \left(TM\otimes T^*M\right) 
        \oplus \Lambda^3T^*M \oplus \Lambda^6T^*M 
        \oplus \Lambda^3TM \oplus \Lambda^6TM . 
\end{aligned}
\end{equation}
The corresponding $\Edd\times\bbR^+$ representations are listed in
Table~\ref{tab:gen-tang}. We write sections as
\begin{equation}
\label{eq:Ed-secs}
\begin{aligned}
   V &= v + \omega + \sigma + \tau && \in E , \\
   Z &= \zeta + u + s + t && \in E^* , \\
   R &= c + r + a + \ta + \alpha + \talpha && \in \adj{\tilde{F}} ,
\end{aligned}
\end{equation}
so that $v\in TM$, $\omega\in\Lambda^2T^*M$, $\zeta\in T^*M$,
$c\in\bbR$ etc. If $\{\hat{e}_a\}$ be a basis for $TM$ with a dual
basis $\{e^a\}$ on $T^*M$ then there is a natural $\gl(d,\bbR)$ action
on each tensor component. For instance 
\begin{equation}
\begin{aligned}
   (r\cdot v)^a &= r^a{}_b v^b , &&&
   (r\cdot \omega)_{ab} 
       &= - r^c{}_a \omega_{cb} - r^c{}_b \omega_{ac} , &&&
   \text{etc}. 
\end{aligned}
\end{equation}

Writing $V'=R\cdot V$ for the adjoint $\Edd\times\bbR^+$ action of
$R\in\adj{\tilde{F}}$ on $V\in E$, the components of $V'$, using the
notation of appendix~\ref{app:tensor}, are given by 
\begin{equation}
\label{eq:E-tranfs}
\begin{aligned}
   v' &= c v + r\cdot v + \alpha\inn\omega - \talpha\inn \sigma  , \\
   \omega' &= c \omega + r\cdot \omega + v\inn a 
         + \alpha\inn \sigma + \talpha\inn\tau , \\
   \sigma' &= c \sigma + r\cdot \sigma + v\inn\ta
         + a\wedge \omega + \alpha\inn \tau , \\
   \tau' &= c \tau + r\cdot \tau 
         - j\ta \wedge \omega + ja\wedge \sigma . 
\end{aligned}
\end{equation}
Note that, the $\Edd$ sub-algebra is generated by setting $c=
\frac{1}{(9-d)} r^a{}_a$. Similarly, given $Z\in E^*$ we have
\begin{equation}
\label{eq:E*-tranfs}
\begin{aligned}
   \zeta' &= - c \zeta + r\cdot \zeta - u\inn a + s\inn \ta  , \\
   u'&= - c u + r\cdot u - \alpha\inn \zeta 
         - s\inn a + t\inn\ta , \\
   s'&= - c s + r\cdot s - \talpha\inn\zeta
         - \alpha\wedge u - t\inn a , \\
   t'&= - c t + r\cdot t
         - j\alpha\wedge s - j\talpha \wedge u . 
\end{aligned}
\end{equation}
Finally the adjoint commutator 
\begin{equation}
   R'' = \BLie{R}{R'} 
\end{equation}
has components
\begin{equation}
\label{eq:adj-tranfs}
\begin{aligned}
   c'' &= \tfrac13 (\alpha \inn a' - \alpha' \inn a) + \tfrac23 (\talpha' \inn \ta - \talpha \inn \ta') , \\
   r'' &= \BLie{r}{r'} + j\alpha \inn ja' - j\alpha' \inn ja - \tfrac13 (\alpha \inn a' - \alpha' \inn a) \id \\
   	& \qquad \qquad + j\talpha' \inn j\ta - j\talpha \inn j\ta' 
	- \tfrac23 (\talpha' \inn \ta - \talpha \inn \ta') \id , \\
   a'' &= r \cdot a' - r' \cdot a + \alpha' \inn \ta - \alpha \inn \ta', \\
   \ta'' &= r \cdot \ta' - r' \cdot \ta - a \wedge a', \\
   \alpha'' &= r \cdot \alpha' - r' \cdot \alpha + \talpha' \inn a - \talpha \inn a' , \\
   \talpha'' &= r \cdot \talpha' - r' \cdot \talpha -\alpha \wedge \alpha'
\end{aligned}
\end{equation}
Here we have $c'' = \tfrac{1}{9-d} r''^a{}_a$, as $R''$ lies in the $\Edd$ sub-algebra.

We also need the projection 
\begin{equation}
   \oadj : E^* \otimes E \to \adj{\tilde{F}} .
\end{equation}
Writing $R=Z\oadj V$ we have 
\begin{equation}
\label{eq:EE*-adj}
\begin{aligned}
	c &= - \tfrac13 u\inn \omega 
		- \tfrac23 s\inn \sigma - t \inn \tau ,\\
	r &= v \otimes \zeta - ju \inn j\omega
		+ \tfrac13 (u\inn\omega) \id 
                - js\inn j\sigma
		+ \tfrac23 (s\inn\sigma) \id
		- jt\inn j\tau ,\\
	\alpha &= v \wedge u +  s\inn\omega + t\inn\sigma ,\\
	\talpha &= - v \wedge s - t\inn\omega , \\
	a &= \zeta \wedge \omega + u\inn \sigma + s\inn \tau ,\\
	\ta &= \zeta \wedge \sigma + u\inn \tau . 
\end{aligned}
\end{equation}
Note in particular that 
\begin{equation}
   \der \oadj V = \der\otimes v + \dd\omega + \dd\sigma . 
\end{equation}


\section{$\Hd$ and $\dHd$}
\label{app:Hd}


We now turn to the analogous description of $\Hd$ in $\SO(d)$
representations. We then give a detailed description of the spinor
representations of $\Hd$ and provide several important projections of
tensor products in this language.


\subsection{$\Hd$ and $\SO(d)$}
\label{app:Hd-Od}

Given a positive definite metric $g$ on $TM$, which for
convenience we take to be in standard form $\delta_{ab}$ in frame
indices, we can define a metric on $E$ by  
\begin{equation}
   G(V,V) = v^2 + \tfrac{1}{2!}\omega^2 
      + \tfrac{1}{5!}\sigma^2 + \tfrac{1}{7!}\tau^2 ,
\end{equation}
where $v^2=v_av^a$, $\omega^2=\omega_{ab}\omega^{ab}$, etc as
in~\eqref{eq:*norm}. Note that this metric allows us to identify
$E\simeq E^*$. 

The subgroup of $\Edd\times\bbR^+$ that leaves the metric is invariant
is $\Hd$, the maximal compact subgroup of $\Edd$ (see
table~\ref{tab:coset}). Geometrically it defines a generalised $\Hd$ 
structure, that is an $\Hd$ sub-bundle $P$ of the generalised
structure bundle $\tilde{F}$. The corresponding Lie algebra bundle is
parametrised by   
\begin{equation}
\label{eq:Hd-alg}
\begin{aligned}
   \adj{P} &\simeq
      \Lambda^2T^*M \oplus \Lambda^3T^*M \oplus \Lambda^6T^*M  , \\
   N &= n + b + \tilde{b} , 
\end{aligned}
\end{equation}
and embeds in $\adj{\tilde{F}}$ as 
\begin{equation}
\label{eq:Hd-embed}
\begin{aligned}
   c &= 0 , \\
   r_{ab} &= n_{ab} , \\
   a_{abc} = - \alpha_{abc} &= b_{abc}, \\
   \tilde{a}_{a_1 \dots a_6} = \tilde{\alpha}_{a_1 \dots a_6} 
      &= \tilde{b}_{a_1 \dots a_6} , 
\end{aligned}
\end{equation}
where indices are lowered with the metric $g$. Note that $n_{ab}$
generates the $O(d)\subset\GL(d,\bbR)$ subgroup that preserves
$g$. Concretely a general group element can be written as 
\begin{equation}
\label{eq:Haction}
   H\cdot V = \ee^{\alpha+\talpha}\,\ee^{a+\ta}\,h\cdot V , 
\end{equation}
where $h\in O(d)$ and $a$ and $\alpha$ and $\ta$ and $\talpha$ are
related as in~\eqref{eq:Hd-embed}. 

The generalised tangent space $E\simeq E^*$ forms an irreducible $\Hd$
bundle, where the action of $\Hd$ just follows
from~\eqref{eq:E-tranfs}. The corresponding representations are listed
in~table~\ref{tab:coset}. 

Another important representation of $\Hd$ is the compliment of the
adjoint of $\Hd$ in $\Edd\times\bbR^+$, which we denote as
$\Hperp$ (see table~\ref{tab:coset}). An element of $\Hperp$ be represented as 
\begin{equation}
\label{eq:H-perp}
\begin{aligned}
   \Hperp &\simeq \bbR \oplus S^2F^* 
      \oplus \Lambda^3F^* \oplus \Lambda^6F^* , \\
   Q &= c + h + q + \tilde{q} 
\end{aligned}
\end{equation}
and it embeds in $\adj{\tilde{F}}$ 
\begin{equation}
\begin{aligned}
   c &= c , \\
   r_{ab} &= h_{ab} , \\
   a_{abc} = \alpha_{abc} &= q_{abc}, \\
   \tilde{a}_{a_1 \dots a_6} = - \tilde{\alpha}_{a_1 \dots a_6} 
      &= \tilde{q}_{a_1 \dots a_6} .
\end{aligned}
\end{equation}
The action of $\Hd$ on this representation is given by the
$\Edd\times\bbR^+$ Lie algebra. Writing $Q' = N\cdot Q$ we have
\begin{equation}
\begin{aligned}
   c' &= -\tfrac23 b\inn q - \tfrac43\tilde{b}\inn\tilde{q} , \\
   h' &= n \cdot h - jb \inn jq - jq \inn jb 
        - j\tilde{b} \inn j\tilde{q} - j\tilde{q} \inn j\tilde{b} 
	+ \big(\tfrac23  b\inn q  + \tfrac43 \tilde{b}\inn\tilde{q} 
              \big) \id , \\
   q' &= n \cdot q - h \cdot b + b \inn \tilde{q} + q \inn \tilde{b} , \\
   \tilde{q}' &= n \cdot \tilde{q} - h \cdot \tilde{b} - b \wedge q ,
\end{aligned}
\end{equation}
where we are using the $\GL(d, \bbR)$ adjoint action of $h$ on
$\Lambda^3T^*M$ and $\Lambda^6T^*M$. The $\Hd$ invariant scalar part
of $Q$ is given by $c-\tfrac{1}{9-d}h^a{}_a$, while the remaining irreducible component 
has $c=\tfrac{1}{9-d}h^a{}_a$.


\subsection{$\dHd$ and $\Cliff(d;\bbR)$}

The double cover $\dHd$ of $\Hd$ has a realisation in terms of the
Clifford algebra $\Cliff(d;\bbR)$. Let $S$ be the bundle of
$\Cliff(d;\bbR)$ spinors. We can identify sections of $S$ as $\dHd$
bundles in two different ways, which we denote $S^\pm$. Specifically
$\chi^\pm\in S^\pm$ if 
\begin{equation}
\label{eq:Hd-cliff}
   N\cdot \chi^\pm = \tfrac{1}{2}\big(\tfrac{1}{2!}n_{ab}\gamma^{ab} 
        \pm \tfrac{1}{3!}b_{abc}\gamma^{abc} 
        - \tfrac{1}{6!}\tb_{a_1\dots a_6}\gamma^{a_1\dots a_6}
        \big) \chi^\pm ,  
\end{equation}
for $N\in\adj{P}$. As expected, in both cases $n$ generates the
$\Spin(d)$ subgroup of $\dHd$. The two representations are mapped into
each other by $\gamma^a \rightarrow -\gamma^a$. As such, they are
inequivalent in odd dimensions. However, in even dimensions, since
$-\gamma^a=\gamma^{(d)} \gamma^a (\gamma^{(d)}){}^{-1}$, they are equivalent
and one can identify $\chi^-=\gamma^{(d)}\chi^+$. Thus one finds 
\begin{equation}
\begin{aligned}
   S &\simeq S^+ \oplus S^- && \text{if $d$ is odd} , \\
   S &\simeq S^+ \simeq S^- && \text{if $d$ is even}.
\end{aligned}
\end{equation}
The different $\dHd$ representations are listed explicitly in
table~\ref{tab:SJ}. 

The $\Spin(d)$ vector-spinor bundle $J$ also forms representations of
$\dHd$. Again we can identify two different actions. If
$\varphi_a^\pm\in J^\pm$ we have\footnote{The formula given here
   matches those found in \cite{Damour:2005zs, de Buyl:2005mt} for
   levels 0, 1 and 2 of $K(E_{10})$. A similar formula also appears in
   the context of $E_{11}$ in~\cite{Steele:2010tk}.} 
\begin{equation}
\label{eq:Hd-action-psi}
\begin{aligned}
   N \cdot \varphi^\pm_a 
      &= \tfrac{1}{2}\big(\tfrac{1}{2!}n_{bc}\gamma^{bc} 
              \pm \tfrac{1}{3!}b_{bcd}\gamma^{bcd} 
              - \tfrac{1}{6!}\tb_{b_1\dots b_6}\gamma^{b_1\dots b_6}
              \big)\varphi^\pm_a 
           - n^b{}_a \varphi^\pm_b 
           \\ & \qquad \qquad 
           \mp \tfrac23 b_a{}^b{}_c \gamma^c \varphi^\pm_b
           \mp \tfrac13 \tfrac{1}{2!} b^b{}_{cd} 
              \gamma_a{}^{cd} \varphi^\pm_b 
           \\ & \qquad \qquad 
           + \tfrac13 \tfrac{1}{4!} \tb_a{}^b{}_{c_1 \dots c_4} 
              \gamma^{c_1 \dots c_4} \varphi^\pm_b
           + \tfrac23 \tfrac{1}{5!} \tb^b{}_{c_1 \dots c_5} 
              \gamma_a{}^{c_1 \dots c_5} \varphi^\pm_b .
\end{aligned}
\end{equation}
Again in even dimension $J^+\simeq J^-$. The $\dHd$ representations
are listed explicitly in table~\ref{tab:SJ}. 

Finally will also need the projections $\Hperp \otimes S^\pm
\rightarrow J^\mp$, which, for $Q\in\Hperp$ and $\chi^\pm \in S^\pm$,
is given by
\begin{equation}
\label{eq:perp-S-proj}
\begin{aligned}
   (Q \proj{J^\mp} \chi^\pm)_a 
      &= \tfrac12 h_{ab} \gamma^b \chi^\pm 
         \mp \tfrac13 \tfrac{1}{2!} q_{abc} \gamma^{bc} \chi^\pm 
         \pm \tfrac16 \tfrac{1}{3!} q_{bcd} 
            \gamma_a{}^{bcd} \chi^\pm 
      \\ & \qquad \qquad 
         + \tfrac16 \tfrac{1}{5!} \tilde{q}_{ab_1 \dots b_5} 
            \gamma^{b_1 \dots b_5} \chi^\pm
         - \tfrac13 \tfrac{1}{6!} \tilde{q}_{c_1 \dots c_6} 
            \gamma_a{}^{c_1 \dots c_6} \chi^\pm.
\end{aligned}
\end{equation}
%


\subsection{$\dHd$ and $\Cliff(10,1;\bbR)$}
\label{app:Hd-cliff101}

To describe the reformulation of $D=11$ supergravity restricted to $d$
dimensions, it is very useful to use the embedding of $\dHd$ in
$\Cliff(10,1;\bbR)$. This identifies the same action of $\dHd$ on
spinors given in~\eqref{eq:Hd-cliff} but now using the internal
spacelike gamma matrices $\Gamma^a$ for $a=1,\dots,d$. Combined with
the external spin generators $\Gamma^{\mu \nu}$, this actually gives
an  action of $\Spin(10-d, 1) \times \dHd$ on eleven-dimensional
spinors. As before the action of $\dHd$ can be embedded in two
different ways. We write $\hat{\chi}^\pm\in\hat{S}^\pm$ with  
\begin{equation}
\label{eq:Hd-cliff101}
   N\cdot\hat{\chi}^\pm 
       = \tfrac{1}{2}\big(\tfrac{1}{2!}n_{ab}\Gamma^{ab} 
             \pm \tfrac{1}{3!}b_{abc}\Gamma^{abc} 
             - \tfrac{1}{6!}\tb_{a_1\dots a_6}\Gamma^{a_1\dots a_6}
          \big) \hat{\chi}^\pm . 
\end{equation}
Since the algebra of the $\{\Gamma^a\}$ is the same as
$\Cliff(d;\bbR)$ all the equations of the previous section translate
directly to this presentation of $\dHd$. The advantage of the direct
action on eleven-dimensional spinors is that it allows us to write
$\dHd$ covariant spinor equations in a dimension independent way. 

As before we can also identify two realisations $\hat{J}^\pm$ of
$\dHd$ on the representations with one eleven-dimensional spinor index
and one internal vector index which transform
as~\eqref{eq:Hd-action-psi} (with $\Gamma^a$ in place of
$\gamma^a$). The $\Spin(d-1,1)\times\dHd$ represenations for
$\hat{S}^\pm$ and $\hat{J}^\pm$ are listed explicitly in
table~\ref{tab:SJodd}. 

In addition to the projection $\Hperp\otimes\hat{S}^\pm\to\hat{J}^\mp$
given by~\eqref{eq:perp-S-proj} (with $\Gamma^a$ in place of
$\gamma^a$) we can identify various other tensor products. We have the
singlet projections $\bl{\cdot}{\cdot}:\hat{S}^\mp \otimes \hat{S}^\pm
\rightarrow \id$ given by the conventional $\Cliff(10,1;\bbR)$
bilinear, defined using~\eqref{eq:11d-conj}, so 
\begin{equation}
\label{eq:Hd-SS-singlet}
   \bl{\hat{\chi}^-}{\hat{\chi}^+} = \bar{\hat{\chi}}^- \chi^+ ,
\end{equation}
where $\hat{\chi}^\pm \in \hat{S}^\pm$. There is a similar singlet
projection $\bl{\cdot}{\cdot}:\hat{J}^\mp \otimes \hat{J}^\pm
\rightarrow \id$ given by\footnote{Setting $d=10$ in this reproduces
   the corresponding inner product in~\cite{Damour:2005zs}.} 
\begin{equation}
\label{eq:Hd-JJ-singlet}
   \bl{\hat{\varphi}^\mp }{\hat{\varphi}^\pm}
      = \bar{\hat{\varphi}}^\mp_a \big(
         \delta^{ab} + \tfrac{1}{9-d} \Gamma^a \Gamma^b 
         \big) \hat{\varphi}^\pm_b ,
\end{equation}
where $\hat{\varphi}^\pm \in \hat{J}^\pm$. 

We also have projections from $\hat{S}^\pm\otimes\hat{J}^\pm$ and
$\hat{S}^\pm \otimes \hat{S}^\mp$ to $\Hperp$. Given
$\hat{\chi}^+\in\hat{S}^-$ and $\hat{\varphi}^\pm\in\hat{J}^\pm$
we have, using the decomposition~\eqref{eq:H-perp}, 
\begin{equation}
\label{eq:SJ-perp-proj}
\begin{aligned}
   (\hat{\chi}^\pm\proj{\Hperp}\hat{\varphi}^\pm)
      &= \tfrac{2}{9-d} \bar{\hat\chi}^\pm \Gamma^a \hat\varphi^\pm_a , \\
   (\hat{\chi}^\pm\proj{\Hperp}\hat{\varphi}^\pm)_{ab}
      &= 2\bar{\hat\chi}^\pm \Gamma^{\ph{\pm}}_{(a} \hat\varphi^\pm_{b)} \\
   (\hat{\chi}^\pm\proj{\Hperp}\hat{\varphi}^\pm)_{abc}
      &= \mp 3\bar{\hat\chi}^\pm \Gamma^{\ph{\pm}}_{[ab} 
         \hat\varphi^\pm_{c]} , \\
   (\hat{\chi}^\pm\proj{\Hperp}\hat{\varphi}^\pm)_{a_1\dots a_6}
      &= -6\bar{\hat\chi}^\pm \Gamma^{\ph{\pm}}_{[a_1 \dots a_5} 
         \hat\varphi^\pm_{a_6]}  ,
\end{aligned}
\end{equation}
Note that the image of this projection does not include the $\dHd$
scalar part of $\Hperp$, since, from the first two components, $c -
\tfrac{1}{9-d} h^a{}_a = 0$. We also have
\begin{equation}
\label{eq:SS-perp-proj}
  (\hat{\chi}^+\proj{\Hperp}\hat{\chi}^-)
      = \tfrac{2}{9-d} \bar{\hat{\chi}}^- \hat{\chi}^+ , 
\end{equation}
and all other components of $\Hperp$ are set to zero. We see that
the image of this map is in the $\dHd$ scalar part of $\Hperp$. 

Finally, we also need the $\dHd$ projections for $E\simeq E^*$ acting
on $S^\pm$ and $J^\pm$. Given $V \in E$ it is useful to introduce the notation 
\begin{equation}
\label{eq:SOd-unique-ops}
\begin{aligned}
   \slashed{\,V} : \hat{S}^\pm &\to \hat{S}^\mp , & &&
   V\Vout : \hat{S}^\pm &\to \hat{J}^\pm , \\
   V\Vin : \hat{J}^\pm &\to \hat{S}^\pm , & &&
   \slashed{\,V}  : \hat{J}^\pm &\to \hat{J}^\mp .
\end{aligned}
\end{equation}
Given $\hat{\chi}^\pm \in \hat{S}^\pm$ and
$\hat{\varphi}^\pm_a \in \hat{J}^\pm$ we find
\begin{align}
\label{eq:SO-1st-proj}
   ( \slashed{\,V} \hat{\chi}^\pm) 
      &= \big(\mp v^a \Gamma_a + \tfrac{1}{2!} \omega_{ab} \Gamma^{ab}
         \mp \tfrac{1}{5!} \sigma_{a_1 \dots a_5} \Gamma^{a_1 \dots a_5}
         + \tfrac{1}{6!} \tau^b{}_{b a_1 \dots a_6} 
            \Gamma^{a_1 \dots a_6}\big) \hat{\chi}^\pm , \\
\intertext{and} 
\label{eq:SO-2nd-proj}
   (V \Vout \hat{\chi}^\pm)_a 
      &= \big( v_a \pm \tfrac23 \Gamma^b \omega_{ab}
         \mp \tfrac13\tfrac{1}{2!}\Gamma_a{}^{cd} \omega_{cd}
         - \tfrac13 \tfrac{1}{4!}  \Gamma^{c_1 \dots c_4} 
            \sigma_{ac_1 \dots c_4} 
         \nonumber \\ & \qquad \qquad
         + \tfrac23\tfrac{1}{5!}  \Gamma_a{}^{c_1 \dots c_5} 
            \sigma_{c_1 \dots c_5}
         \pm \tfrac{1}{7!}  \Gamma^{c_1 \dots c_7} 
            \tau_{a,c_1 \dots c_7} \big) \hat{\chi}^\pm , 
\intertext{while}
\label{eq:SO-3rd-proj}
   (V \Vin \hat{\varphi}^\pm) 
      &= v^a \hat{\varphi}^\pm_a 
         + \tfrac{1}{10-d} v_a \Gamma^{ab} \hat{\varphi}^\pm_b 
         \pm \tfrac{1}{10-d} \tfrac{1}{2!} \omega_{bc} 
            \Gamma^{abc} \hat{\varphi}^\pm_a 
         \pm \tfrac{8-d}{10-d} \omega^a{}_b \Gamma^b \hat{\varphi}^\pm_a 
         \nonumber \\ & \qquad \qquad
        - \tfrac{1}{10-d} \tfrac{1}{5!} \sigma^{b_1 \dots b_5} 
            \Gamma^a{}_{b_1 \dots b_5} \hat{\varphi}^\pm_a
         - \tfrac{8-d}{10-d} \tfrac{1}{4!} \sigma^a{}_{b_1 \dots b_4} 
            \Gamma^{b_1 \dots b_4} \hat{\varphi}^\pm_a 
         \nonumber \\ & \qquad \qquad
        \mp\tfrac{1}{7!} \tau^a{}_{,b_1 \dots b_7} 
            \Gamma^{b_1 \dots b_7} \hat{\varphi}^\pm_a
         \mp\tfrac13 \tfrac{1}{5!} \tau^c{}_{,c}{}^a{}_{b_1 \dots b_5} 
            \Gamma^{b_1 \dots b_5} \hat{\varphi}^\pm_a ,
\intertext{and finally}
\label{eq:SO-4th-proj}
   ( \slashed{\,V} \hat{\varphi}^\pm)_a 
      &= \pm v^c \Gamma_c \hat{\varphi}^\pm_a 
         \pm \tfrac{2}{9-d} \Gamma^c v_a \hat{\varphi}^\pm_c  
         - \tfrac{1}{2!} \omega_{cd} \Gamma^{cd} \hat{\varphi}^\pm_a 
         + \tfrac43 \omega_a{}^b \hat{\varphi}^\pm_b 
         \nonumber \\ & \qquad
         - \tfrac23 \omega_{cd} \Gamma_a{}^c \hat{\varphi}^{\pm d} 
         - \tfrac43 \tfrac{1}{9-d} \omega_{ab} 
             \Gamma^b\Gamma^c \hat{\varphi}^\pm_c
         + \tfrac23 \tfrac{1}{9-d} \tfrac{1}{2!} \omega_{bc}
            \Gamma_a{}^{bc} \Gamma^d \hat{\varphi}^\pm_d 
         \nonumber \\ & \qquad
         \pm \tfrac{1}{5!} \sigma_{c_1 \dots c_5} \Gamma^{c_1 \dots c_5} 
            \hat{\varphi}^\pm_a
         \mp \tfrac23 \tfrac{1}{3!}\sigma_a{}^b{}_{c_1 c_2 c_3} 
            \Gamma^{c_1 c_2 c_3} \hat{\varphi}^\pm_b
         \mp \tfrac43 \tfrac{1}{4!}\sigma^b{}_{c_1 \dots c_4} 
            \Gamma_a{}^{c_1 \dots c_4} \hat{\varphi}^\pm_b 
         \nonumber \\ & \qquad
         \mp \tfrac23 \tfrac{1}{9-d} \tfrac{1}{4!}
            \sigma_{ac_1 \dots c_4} \Gamma^{c_1 \dots c_4} 
            \Gamma^d \hat{\varphi}^\pm_d
         \pm \tfrac43 \tfrac{1}{9-d} \tfrac{1}{5!}
            \sigma_{c_1 \dots c_5} \Gamma_a{}^{c_1 \dots c_5} 
            \Gamma^d \hat{\varphi}^\pm_d 
         \nonumber \\ & \qquad
         + \tfrac{1}{7!} \tau_{c,d_1 \dots d_7} 
            \Gamma^c \Gamma^{d_1 \dots d_7} \hat{\varphi}^\pm_a
         + \tfrac{1}{7!} \tau_{a,c_1 \dots c_7} 
            \Gamma^{c_1 \dots c_7} \Gamma^d \hat{\varphi}^\pm_d .
\end{align}
%




\begin{thebibliography}{99}



\bibitem{CSW1}
   A.~Coimbra, C.~Strickland-Constable, D.~Waldram,
   ``Supergravity as Generalised Geometry I: Type II Theories,''
  JHEP {\bf 1111} (2011) 091
  [arXiv:1107.1733 [hep-th]].
\bibitem{GCY}
   N.~Hitchin,
   ``Generalized Calabi-Yau manifolds,''
   Quart.\ J.\ Math.\ Oxford Ser.\  {\bf 54}, 281 (2003) 
   [arXiv:math.dg/0209099].
\bibitem{Gualtieri}
   M.~Gualtieri, 
   ``Generalized Complex Geometry,''
   Oxford University DPhil thesis (2004) [arXiv:math.DG/0401221] 
   and [arXiv:math.DG/0703298].
\bibitem{double1}
   O.~Hohm, C.~Hull and B.~Zwiebach,
   ``Background independent action for double field theory,''
   JHEP {\bf 1007}, 016 (2010)
   [arXiv:1003.5027 [hep-th]]; \\
   O.~Hohm, C.~Hull and B.~Zwiebach,
   ``Generalized metric formulation of double field theory,''
   JHEP {\bf 1008}, 008 (2010)
   [arXiv:1006.4823 [hep-th]]; \\
   O.~Hohm and S.~K.~Kwak,
   ``Frame-like Geometry of Double Field Theory,''
   J.\ Phys.\ A {\bf 44}, 085404 (2011)
   [arXiv:1011.4101 [hep-th]].
\bibitem{siegel}
   W.~Siegel,
   ``Two vierbein formalism for string inspired axionic gravity,''
   Phys.\ Rev.\  {\bf D47 } (1993)  5453-5459.
   [hep-th/9302036], \\
   W.~Siegel,
   ``Superspace duality in low-energy superstrings,''
   Phys.\ Rev.\  {\bf D48 } (1993)  2826-2837.
   [hep-th/9305073].
\bibitem{double2}
   O.~Hohm, S.~K.~Kwak and B.~Zwiebach,
   ``Unification of Type II Strings and T-duality,''
   Phys.\ Rev.\ Lett.\  {\bf 107}, 171603 (2011)
   [arXiv:1106.5452 [hep-th]]; \\
   O.~Hohm, S.~K.~Kwak and B.~Zwiebach,
   ``Double Field Theory of Type II Strings,''
   JHEP {\bf 1109}, 013 (2011)
   [arXiv:1107.0008 [hep-th]]. 
\bibitem{double-geom}
   O.~Hohm and B.~Zwiebach,
   ``Large Gauge Transformations in Double Field Theory,''
   JHEP {\bf 1302}, 075 (2013)
   [arXiv:1207.4198 [hep-th]].
\bibitem{CSW2} 
   A.~Coimbra, C.~Strickland-Constable and D.~Waldram,
   ``$E_{d(d)} \times \mathbb{R}^+$ Generalised Geometry, Connections
   and M theory,'' 
   arXiv:1112.3989 [hep-th].
\bibitem{JLP}
   I.~Jeon, K.~Lee, J.-H.~Park,
   ``Differential geometry with a projection: Application to double
   field theory,'' 
   JHEP {\bf 1104 } (2011)  014.
   [arXiv:1011.1324 [hep-th]]; \\
   I.~Jeon, K.~Lee, J.~-H.~Park,
   ``Stringy differential geometry, beyond Riemann,''  
   [arXiv:1105.6294 [hep-th]].
\bibitem{chris}
   C.~M.~Hull,
   ``Generalised Geometry for M-Theory,''
   JHEP {\bf 0707}, 079 (2007)
   [arXiv:hep-th/0701203].
\bibitem{PW}
   P.~P.~Pacheco, D.~Waldram,
   ``M-theory, exceptional generalised geometry and superpotentials,''
   JHEP {\bf 0809}, 123 (2008).
   [arXiv:0804.1362 [hep-th]].
\bibitem{duff2}
   M.~J.~Duff,
   ``$E_8\times\SO(16)$ Symmetry of $d=11$ Supergravity,''
   in \textit{Quantum field theory and quantum statistics}, vol. 2,
   p209, eds. I.~A.~Batalin et al., Adam Hilger (1987) (CERN-TH-4124). 
\bibitem{deWN}
   B.~de Wit and H.~Nicolai,
   ``$D = 11$ Supergravity With Local $\SU(8)$ Invariance,''
   Nucl.\ Phys.\ B {\bf 274}, 363 (1986).
\bibitem{nic}
   H.~Nicolai,
   ``$D = 11$ Supergravity with Local $\SO(16)$ Invariance,''
   Phys.\ Lett.\  B {\bf 187}, 316 (1987).
\bibitem{west} 
   P.~C.~West,
   ``Hidden superconformal symmetry in M theory,''
   JHEP {\bf 0008}, 007 (2000)
   [hep-th/0005270]; \\
   P.~C.~West,
   ``$E_{11}$ and M theory,''
   Class.\ Quant.\ Grav.\ \ {\bf 18}, 4443  (2001)
   [hep-th/0104081]; \\
   P.~C.~West,
   ``E(11), SL(32) and central charges,''
   Phys.\ Lett.\ B {\bf 575}, 333 (2003)
   [hep-th/0307098].
\bibitem{KNS}
   K.~Koepsell, H.~Nicolai and H.~Samtleben,
   ``An exceptional geometry for $d=11$ supergravity?,''
   Class.\ Quant.\ Grav.\  {\bf 17}, 3689 (2000)
   [arXiv:hep-th/0006034].
\bibitem{dewitBPS}
   B.~de Wit,
   ``M theory duality and BPS extended supergravity,''
   Int.\ J.\ Mod.\ Phys.\ A {\bf 16}, 1002 (2001)
   [hep-th/0010292].
\bibitem{deWN-extra}
   B.~de Wit and H.~Nicolai,
   ``Hidden symmetries, central charges and all that,''
   Class.\ Quant.\ Grav.\ \ {\bf 18}, 3095  (2001)
   [hep-th/0011239].
\bibitem{E10}
   T.~Damour, M.~Henneaux and H.~Nicolai,
   ``$E_{10}$ and a 'small tension expansion' of M theory,''
   Phys.\ Rev.\ Lett.\  {\bf 89}, 221601 (2002)
   [hep-th/0207267].
   T.~Damour, M.~Henneaux and H.~Nicolai,
   ``Cosmological billiards,''
   Class.\ Quant.\ Grav.\  {\bf 20}, R145 (2003)
   [hep-th/0212256].
\bibitem{hillmann}
   C.~Hillmann,
   ``Generalized $E_{7(7)}$ coset dynamics and $D=11$ supergravity,''
   JHEP {\bf 0903}, 135 (2009).
   [arXiv:0901.1581 [hep-th]], \\
   C.~Hillmann,
   ``$E_{7(7)}$ and $d=11$ supergravity,''
   [arXiv:0902.1509 [hep-th]].
\bibitem{BP}
   D.~S.~Berman and M.~J.~Perry,
   ``Generalized Geometry and M theory,''
   JHEP\ {\bf 1106}, 074  (2011)
   [arXiv:1008.1763 [hep-th]]; \\
   D.~S.~Berman, H.~Godazgar and M.~J.~Perry,
   ``$SO(5,5)$ duality in M-theory and generalized geometry,''
   Phys.\ Lett.\ B\ {\bf 700}, 65  (2011)
   [arXiv:1103.5733 [hep-th]]; \\
   D.~S.~Berman, H.~Godazgar, M.~J.~Perry and P.~West,
   ``Duality Invariant Actions and Generalised Geometry,''
   JHEP {\bf 1202}, 108 (2012)
   [arXiv:1111.0459 [hep-th]].
\bibitem{GLSW}
   M.~Grana, J.~Louis, A.~Sim, D.~Waldram,
   ``$E_{7(7)}$ formulation of $N=2$ backgrounds,''
   JHEP {\bf 0907}, 104 (2009).
   [arXiv:0904.2333 [hep-th]].
\bibitem{E7-flux}
   G.~Aldazabal, E.~Andres, P.~G.~Camara and M.~Grana,
   ``U-dual fluxes and Generalized Geometry,''
   JHEP\ {\bf 1011}, 083  (2010)
   [arXiv:1007.5509 [hep-th]].
\bibitem{GO1}
   M.~Grana and F.~Orsi,
   ``$N=1$ vacua in Exceptional Generalized Geometry,''
   JHEP {\bf 1108}, 109 (2011)
   [arXiv:1105.4855 [hep-th]].
\bibitem{Thompson:2011uw} 
   D.~C.~Thompson,
   ``Duality Invariance: From M-theory to Double Field Theory,''
   JHEP {\bf 1108}, 125 (2011)
   [arXiv:1106.4036 [hep-th]].
\bibitem{BMJ}
   D.~S.~Berman, E.~T.~Musaev and M.~J.~Perry,
   ``Boundary Terms in Generalized Geometry and doubled field theory,''
   Phys.\ Lett.\ B {\bf 706}, 228 (2011)
   [arXiv:1110.3097 [hep-th]].
\bibitem{BP-alg}
   D.~S.~Berman, H.~Godazgar, M.~Godazgar and M.~J.~Perry,
   ``The Local symmetries of M-theory and their formulation in
   generalised geometry,'' 
   JHEP {\bf 1201}, 012 (2012)
   [arXiv:1110.3930 [hep-th]].
\bibitem{Cassani:2011fu} 
   D.~Cassani and P.~Koerber,
   ``Tri-Sasakian consistent reduction,''
   JHEP {\bf 1201}, 086 (2012)
   [arXiv:1110.5327 [hep-th]].
\bibitem{GO2}
   M.~Grana and F.~Orsi,
   ``$N=2$ vacua in Generalized Geometry,''
   JHEP {\bf 1211}, 052 (2012)
   [arXiv:1207.3004 [hep-th]].
\bibitem{BMT}
   D.~S.~Berman, E.~T.~Musaev and D.~C.~Thompson,
   ``Duality Invariant M-theory: Gauged supergravities and
   Scherk-Schwarz reductions,'' 
   JHEP {\bf 1210}, 174 (2012)
   [arXiv:1208.0020 [hep-th]].
\bibitem{BCKT}
   D.~S.~Berman, M.~Cederwall, A.~Kleinschmidt and D.~C.~Thompson,
   ``The gauge structure of generalised diffeomorphisms,''
   arXiv:1208.5884 [hep-th].
\bibitem{GT} 
   M.~Grana and H.~Triendl,
   ``Generalized $N=1$ and $N=2$ structures in M-theory and type II
   orientifolds,'' 
   arXiv:1211.3867 [hep-th].



\bibitem{duff1}
   M.~J.~Duff,
   ``Duality Rotations In String Theory,''
   Nucl.\ Phys.\ B {\bf 335}, 610 (1990); \\
   M.~J.~Duff and J.~X.~Lu,
   ``Duality Rotations In Membrane Theory,''
   Nucl.\ Phys.\ B {\bf 347}, 394 (1990).
\bibitem{dft}
   C.~Hull, B.~Zwiebach,
   ``Double Field Theory,''
   JHEP {\bf 0909 } (2009)  099.
   [arXiv:0904.4664 [hep-th]].
\bibitem{BO}
   A.~B.~Borisov and V.~I.~Ogievetsky,
   ``Theory of Dynamical Affine and Conformal Symmetries as Gravity
   Theory,'' 
   Theor.\ Math.\ Phys.\ \ {\bf 21}, 1179  (1975)
   [Teor.\ Mat.\ Fiz.\ \ {\bf 21}, 329  (1974)].

\bibitem{GP}
   J.~P.~Gauntlett, S.~Pakis,
   ``The Geometry of D = 11 killing spinors,''
   JHEP {\bf 0304}, 039 (2003).
   [hep-th/0212008].
\bibitem{gerbes}
   N.~J.~Hitchin,
   ``Lectures on special Lagrangian submanifolds,''
   arXiv:math/9907034.
\bibitem{embedT1}
  B.~de Wit, H.~Samtleben, M.~Trigiante,
  ``On Lagrangians and gaugings of maximal supergravities,''
  Nucl.\ Phys.\  {\bf B655}, 93-126 (2003).
  [hep-th/0212239]; \\
  B.~de Wit, H.~Samtleben, M.~Trigiante,
  ``The Maximal D=4 supergravities,''
  JHEP {\bf 0706}, 049 (2007).
  [arXiv:0705.2101 [hep-th]].
\bibitem{embedT2}
  A.~Le Diffon, H.~Samtleben, M.~Trigiante,
  ``N=8 Supergravity with Local Scaling Symmetry,''
  JHEP {\bf 1104}, 079 (2011).
  [arXiv:1103.2785 [hep-th]].
\bibitem{trombone}
  E.~Cremmer, H.~Lu, C.~N.~Pope, K.~S.~Stelle,
  ``Spectrum generating symmetries for BPS solitons,''
  Nucl.\ Phys.\  {\bf B520}, 132-156 (1998).
  [hep-th/9707207].
\bibitem{GMPW}
   M.~Grana, R.~Minasian, M.~Petrini and D.~Waldram,
   ``T-duality, Generalized Geometry and Non-Geometric Backgrounds,''
   JHEP {\bf 0904}, 075 (2009)
   [arXiv:0807.4527 [hep-th]].
\bibitem{keur}
   A.~Keurentjes,
   ``U duality (sub)groups and their topology,''
   Class.\ Quant.\ Grav.\  {\bf 21}, S1367 (2004)
   [hep-th/0312134].
  
\bibitem{gen-holo-duff} 
   M.~J.~Duff and J.~T.~Liu,
   ``Hidden space-time symmetries and generalized holonomy in M theory,''
   Nucl.\ Phys.\ B {\bf 674}, 217 (2003)
   [hep-th/0303140],
\bibitem{sara}
  A.~Coimbra, Imperial College PhD transfer report (2010), \\
  S.~Tavares, Imperial College MSc thesis (2010).
\bibitem{Damour:2005zs} 
  T.~Damour, A.~Kleinschmidt and H.~Nicolai,
  ``Hidden symmetries and the fermionic sector of eleven-dimensional supergravity,''
  Phys.\ Lett.\ B {\bf 634}, 319 (2006)
  [hep-th/0512163].
\bibitem{de Buyl:2005mt} 
  S.~de Buyl, M.~Henneaux and L.~Paulot,
  ``Extended E(8) invariance of 11-dimensional supergravity,''
  JHEP {\bf 0602}, 056 (2006)
  [hep-th/0512292].
\bibitem{Steele:2010tk} 
  D.~Steele and P.~West,
  ``E11 and Supersymmetry,''
  JHEP {\bf 1102}, 101 (2011)
  [arXiv:1011.5820 [hep-th]].
 

\end{thebibliography}
\end{document}